\documentclass[journal,a4paper]{IEEEtran}
\usepackage{booktabs}
\usepackage{graphicx} 
\usepackage[hyphens]{url}  
\usepackage{algorithm}
\usepackage{algorithmic}
\usepackage{listings}
\usepackage{comment}
\usepackage{color}

\usepackage{amsmath}
\usepackage{amsfonts}
\usepackage{multicol}
\usepackage{multirow}
\usepackage{dirtytalk}
\usepackage{soul}

\floatstyle{ruled}
\newfloat{listing}{tb}{lst}{}
\floatname{listing}{Listing}

\usepackage{amsmath,amsfonts}
\usepackage{algorithmic}
\usepackage{algorithm}
\usepackage{array}
\usepackage[caption=false]{subfig}
\usepackage{textcomp}
\usepackage{stfloats}
\usepackage{url}
\usepackage{verbatim}
\usepackage{graphicx}
\def\BibTeX{{\rm B\kern-.05em{\sc i\kern-.025em b}\kern-.08em
    T\kern-.1667em\lower.7ex\hbox{E}\kern-.125emX}}
\begin{document}

\title{Cross-Attention Message-Passing Transformers for Code-Agnostic Decoding in 6G Networks}

\author{Seong-Joon~Park,~\IEEEmembership{Member,~IEEE,}
Hee-Youl~Kwak,~\IEEEmembership{Member,~IEEE,}
Sang-Hyo~Kim,~\IEEEmembership{Member,~IEEE,}
Yongjune~Kim,~\IEEEmembership{Member,~IEEE,}
Jong-Seon~No,~\IEEEmembership{Fellow,~IEEE}


\thanks{A preliminary version of this work was presented in part at the Thirteenth International Conference on Learning Representations (ICLR), Singapore, April 2025 \cite{Park2025crossmpt}. (Corresponding authors: H.-Y. Kwak and Y. Kim.)}
\thanks{Seong-Joon Park is with the Institute of Artificial Intelligence, Pohang University of Science and Technology (POSTECH), Pohang 37673, South Korea (e-mail: joonpark2247@gmail.com).}
\thanks{Hee-Youl Kwak is with the Department of Electrical, Electronic and Computer Engineering, University of Ulsan, Ulsan 44610, South Korea (e-mail: ghy1228@gmail.com).}
\thanks{Sang-Hyo Kim is with the Department of Electrical and Computer Engineering, Sungkyunkwan University, Suwon 16419, South Korea (e-mail: iamshkim@skku.edu).}
\thanks{Yongjune Kim is with the Department of Electrical Engineering, Pohang University of Science and Technology (POSTECH), Pohang 37673, South Korea (e-mail: yongjune@postech.ac.kr).}
\thanks{Jong-Seon No is with the Department of Electrical and Computer Engineering, Seoul National University, Seoul 08826, South Korea (e-mail: jsno@snu.ac.kr).}

}



\maketitle

\begin{abstract}
Channel coding for 6G networks is expected to support a wide range of requirements arising from heterogeneous communication scenarios.
These demands challenge traditional code-specific decoders, which lack the flexibility and scalability required for next-generation systems.
To tackle this problem, we propose an AI-native foundation model for unified and code-agnostic decoding based on the transformer architecture.
We first introduce a cross-attention message-passing transformer~(CrossMPT).
CrossMPT employs two masked cross-attention blocks that iteratively update two distinct input representations--magnitude and syndrome vectors--allowing the model to effectively learn the decoding problem.
Notably, our CrossMPT has achieved state-of-the-art decoding performance among single neural decoders.
Building on this, we develop foundation CrossMPT~(FCrossMPT) by making the architecture invariant to code length, rate, and class, allowing a single trained model to decode a broad range of codes without retraining.
To further enhance decoding performance, particularly for short blocklength codes, we propose CrossMPT ensemble decoder~(CrossED), an ensemble decoder composed of multiple parallel CrossMPT blocks employing different parity-check matrices.
This architecture can also serve as a foundation model, showing strong generalization across diverse code types.
Overall, the proposed AI-native code-agnostic decoder offers flexibility, scalability, and high performance, presenting a promising direction to channel coding for 6G networks.

\end{abstract}


\section{Introduction}

\IEEEPARstart{T}{he} sixth generation (6G) of wireless communication is envisioned to support an unprecedented range of scenarios, from hyper-reliable low-latency communication (HRLLC) to massive communication~\cite{b_Saad2020, b_Jiang2021}.
This expands upon the 5G standard, which already employs different coding schemes like low-density parity-check~(LDPC) and polar codes for different use cases~\cite{b_Miao2024, b_Geiselhart2023}.
Continuing this trend for each diverse 6G scenario would require a growing number of specialized codes and their corresponding decoders, which complicates system integration and increases hardware overhead~\cite{b_Geiselhart2023}.
This broad range of scenarios imposes diverse requirements on the physical layer, particularly on channel coding, and calls for a highly flexible and adaptable decoding approach~\cite{b_Geiselhart2023,b_Zhang2023}.  


To meet these challenges, a unified and code-agnostic decoder capable of handling various code types, lengths, and rates is essential for enabling flexible and scalable channel coding in 6G networks~\cite{b_Miao2024,b_Geiselhart2023}. However, most decoders are designed for specific codes with fixed parameters, limiting their flexibility~\cite{b_Geiselhart2023}. While some algorithms can decode multiple code classes and are considered universal decoders, they still require separate implementations for different code structures.

This limitation highlights the need for a foundation model in error correcting code~(ECC) decoding; one that can generalize across diverse codes without requiring code-specific adaptation.
Enabled by recent advances in deep learning techniques~\cite{b_bert,b_resnet,b_od1,b_detr}, AI-based ECC decoders can learn complex representations and generalize across a wide range of code structures, paving the way for a foundation model for ECC decoding.

\subsection{Main Contributions}
In this paper, we propose an AI-native, unified, and code-agnostic decoder using the transformer architecture.
We first propose the cross-attention message-passing transformer~(CrossMPT) that employs cross-attention modules to emulate the message-passing decoding algorithm.
Our CrossMPT effectively updates the magnitude and syndrome of received vectors iteratively using two masked cross-attention blocks.
These blocks utilize the parity-check matrix (PCM) and its transpose as attention masks, guiding the model to focus exclusively on connected node interactions defined by the code structure.
As in traditional message-passing decoders such as belief propagation (BP)~\cite{b_BP} or min-sum~(MS) algorithm~\cite{b_MS}, where messages are exchanged only between connected variable and check nodes, the masked cross-attention mechanism ensures that the attention scores are computed only over valid edges specified by the PCM.
By integrating the transformer's powerful attention mechanism into the message-passing decoding principle, our CrossMPT has achieved state-of-the-art decoding performance among single neural decoders.

Furthermore, we propose a foundation model of CrossMPT~(FCrossMPT) by replacing the code-dependent parameters of CrossMPT with code-invariant embeddings.
This design makes the model fully agnostic to code-specific properties such as code class, length, and rate, enabling it to be trained on diverse codes with varying configurations.
As a result, a single trained model can effectively decode multiple types of codes without retraining or architectural modifications.

To boost the decoding performance of short blocklength codes, we propose CrossMPT ensemble decoder~(CrossED)--an ensemble decoding architecture that consists of multiple cross-attention modules, each employing a distinct attention mask derived from a different PCM.
Ensemble decoding is well-aligned with the requirements of 6G networks, particularly HRLLC, as it enhances decoding accuracy through diversity while maintaining low decoding latency through parallel processing.
This is especially critical in short blocklength regimes, where conventional decoders often face performance limitations under strict latency constraints.

Our ensemble design leverages complementary attention masks generated by cyclically shifting the systematic PCM, allowing each decoder to attend to different structural aspects of the code.
These complementary masks introduce decoding diversity by providing multiple perspectives on bit-level dependencies within the same codeword.
As a result, the ensemble design can compensate for decoding failures that may occur in any single masking configuration.
Since all decoder modules operate in parallel, the system can achieve performance gains without incurring additional latency.

Leveraging recent advances in AI model, especially the transformer architecture, we present the following novel contributions and results that align well with the requirements and challenges of 6G networks:
\begin{enumerate}
    \item We propose a novel transformer-based decoder called CrossMPT, which shares key operational principles with conventional message-passing decoders. 
    CrossMPT outperforms the original transformer-based decoder in both decoding performance and computational complexity, and has achieved state-of-the-art decoding performance among single neural decoders.
    
    \item Building upon CrossMPT, we propose an AI-native foundation model for ECC decoding, FCrossMPT, which can be trained not only on a wide range of code lengths and code rates, but also on different code classes within a single model.
    We demonstrate that a single trained model can effectively decode diverse codes without requiring separate implementations or weight adjustments.
    This approach establishes a unified, code-agnostic decoder that serves as a true foundation model for ECC decoding.
    

    \item To enhance the decoding performance of short-length codes, we propose CrossED, an ensemble decoder composed of multiple CrossMPT blocks in a parallel manner.
    Each CrossMPT block employs distinct PCMs as mask matrices, enabling the ensemble to capture diverse code structures. 
    This architecture significantly improves decoding performance while maintaining decoding latency.
    Moreover, our CrossED is designed as a foundation model for ECC decoding, capable of decoding multiple code types without code-specific adaptations.
       
\end{enumerate}

\subsection{Related Work}
The diverse requirements of 6G networks have made a unified, code-agnostic decoding scheme a key objective \cite{b_Miao2024, b_Geiselhart2023} and this has led to strong interest in universal decoders. Classical algorithms like ordered statistics decoding (OSD)~\cite{b_fossorier1995_osd} and guessing random additive noise decoding (GRAND)~\cite{b_duffy2019_grand} are considered universal as they can be applied to various code classes \cite{b_Geiselhart2023}. Similarly, our transformer-based decoder can function as a universal decoder; furthermore, it can be extended to become a foundation model--a single, pre-trained network that generalizes across diverse codes without requiring code-specific parameters or additional training.

In addition, ensemble decoding is seen as a crucial strategy for 6G, as it enhances reliability for short blocklength codes by running multiple decoders in parallel, without increasing the overall latency \cite{b_Geiselhart2023}.
Our proposed CrossED architecture implements this principle to effectively boost the decoding performance using parallel neural networks.

In this context, we briefly review deep learning approaches to ECC decoding, which can be broadly classified into two categories: \emph{model-based} and \emph{model-free}.

\subsubsection{Model-Based Approach}
The first approach is to implement conventional decoding methods, such as the BP decoder and min-sum~(MS) decoder, within a neural network framework.
These neural decoders unfold the iterative decoding process on the Tanner graph into a deep neural network.
Nachmani {\textit{et al.}}~\cite{b_Nachmani2018} proposed a neural decoder based on a recurrent neural network for BCH codes, achieving improved decoding performance over the standard BP decoder.
Dai {\textit{et al.}}~\cite{b_Dai2021} modified the neural MS decoder for protograph LDPC codes, introducing a parameter-sharing mechanism that enhances scalability for long codes. 
Furthermore, several studies exhibited that neural network-based BP and MS decoders can outperform the traditional decoders~\cite{b_Nachmani2019,b_Nachmani2021,b_Kwak2024, b_Kwak2022,b_Kwak2025, b_Buchberger2021}.
However, these model-based neural decoders might face restrictive performance limits due to architectural constraints. 

\subsubsection{Model-Free Approach}

The second approach is a model-free approach, which employs neural network architectures without relying on prior knowledge of conventional decoding algorithms.
Early attempts using fully connected networks were often hampered by significant overfitting issues during training~\cite{b_Gruber2017, b_Cammerer2017}. 
This approach is not restricted to the conventional decoding models but initially encounters the overfitting problem, as it is impractical to train on all possible codewords.
To address this, Bennatan {\textit{et al.}}~\cite{b_Bennatan2018} proposed a preprocessing that enables model-free decoders to overcome the overfitting problem~\cite{b_ECCT}.
They leveraged the syndrome to learn noise using only all-zero codeword and incorporated a recurrent neural network architecture with this preprocessing step.

Building on this, Choukroun and Wolf advanced this concept with the Error Correction Code Transformer (ECCT)~\cite{b_ECCT}, which avoids overfitting through the same preprocessing method and achieves state-of-the-art decoding performance by utilizing a masked self-attention mechanism.
The ECCT framework has since inspired further research; for example, denoising diffusion ECCs~\cite{b_DDECC} re-envisioned decoding as a diffusion process, while multiple-masks ECCT~\cite{Park2024multiple} employed varied PCMs for the same code to enhance performance by capturing diverse bit relationships.

Furthermore, Choukroun and Wolf introduced the foundation ECCT (FECCT)~\cite{b_Choukroun2024}, the first foundation model for transformer-based decoding, designed to be agnostic to code class, length, and rate.
While FECCT enhances generalization performance through sophisticated learned mappings, resulting dense attention mechanism leads to high computational load and complexity.
In contrast, our proposed FCrossMPT also functions as a foundation model but retains a simple and sparse masking structure derived directly from the PCM and its transpose.
This simple structure allows FCrossMPT to achieve broad generalization while maintaining significantly lower computational complexity than FECCT.


\section{Background}
\label{sec_background}
In this section, we provide a concise overview of the background of the ECCs and transformer architecture, along with the preprocessing and postprocessing techniques used in the original ECCT~\cite{b_ECCT}.

\subsection{Error Correcting Codes}

A linear block code defines a set of codewords $C \subset \{0,1\}^n$ generated as linear combinations of the rows of a generator matrix $\mathbf{G} \in \mathbb{F}_2^{k \times n}$. To verify whether a received vector belongs to the code, a PCM $\mathbf{H} \in \mathbb{F}_2^{(n-k) \times n}$ is used, which satisfies $\mathbf{G}\mathbf{H}^\top = 0$ over $GF(2)$. A vector $\mathbf{x} \in \{0,1\}^n$ is a valid codeword if and only if $\mathbf{H}\mathbf{x} = 0$. Notably, multiple valid PCMs can represent the same code, as elementary row operations on $\mathbf{H}$ preserve its null space.

When transmitting over an additive white Gaussian noise (AWGN) channel, a binary codeword $\mathbf{x}$ is modulated using binary phase shift keying (BPSK) as $\mathbf{x}_s$, and the received signal is modeled as $\mathbf{y} = \mathbf{x}_s + \mathbf{z}$, where $\mathbf{z} \sim \mathcal{N}(0, \sigma^2)$ is i.i.d. Gaussian noise. The receiver attempts to recover $\mathbf{x}$ by first computing a hard-decision vector $\mathbf{y}_b = \mathrm{bin}(\mathrm{sign}(\mathbf{y}))$ and then calculating the syndrome $s(\mathbf{y}) = \mathbf{H}\mathbf{y}_b$. If $s(\mathbf{y}) \neq 0$, the decoder recognizes that errors are present and proceeds with error correction.

\subsection{Transformer Architecture}

The transformer architecture, initially developed for natural language processing~\cite{b_transformer}, has become one of the most widely adopted neural network models across diverse domains.
Its superior modeling capabilities and scalability have made it a dominant architecture in modern deep learning, consistently achieving state-of-the-art performance on various tasks.

The transformer architecture, first introduced for natural language processing (NLP), has demonstrated remarkable versatility and is now one of the most widely adopted neural network models.
Its application has expanded to diverse domains including computer vision, speech recognition, and even computational biology~\cite{b_vit, b_dong2018_speech, b_jumper2021}.
Its superior modeling capabilities and scalability have made it a dominant architecture in modern deep learning, consistently achieving state-of-the-art performance on various tasks.


A core component of the transformer architecture is the \textit{attention module}, which enables the model to dynamically focus on relevant parts of the input sequence based on learned dependencies. It operates on three key elements: the query ($\mathbf{Q}$), key ($\mathbf{K}$), and value ($\mathbf{V}$) matrices. Given two input embeddings $\mathbf{X}, \mathbf{X'} \in \mathbb{R}^{n \times d}$, where $n$ is the sequence length and $d$ is the embedding dimension, these matrices are linearly projected as
$\mathbf{Q} = \mathbf{X} \mathbf{W}_Q, \quad \mathbf{K} = \mathbf{X'} \mathbf{W}_K, \quad \mathbf{V} = \mathbf{X'} \mathbf{W}_V,$
where $\mathbf{W}_Q, \mathbf{W}_K, \mathbf{W}_V \in \mathbb{R}^{d \times d}$ are trainable parameters.

The attention output is then computed using the scaled dot-product attention mechanism:
\begin{equation*}
 \mathrm{Attention}(\mathbf{Q}, \mathbf{K}, \mathbf{V}) = \mathrm{softmax} \left( \frac{\mathbf{Q} \mathbf{K}^\top}{\sqrt{d}} \right) \mathbf{V}.   
\end{equation*}
This operation yields a weighted sum of the value vectors, with the weights determined by the similarity between the queries and keys. The scaling factor $\sqrt{d}$ is introduced for numerical stability.
If the same input embedding is used for both the query and key/value ($\mathbf{X} = \mathbf{X'}$), the operation is referred to as \textit{self-attention}. When the query and key/value come from different sources ($\mathbf{X} \neq \mathbf{X'}$), the operation is known as \textit{cross-attention}.

\begin{figure}[!t]
\begin{center}
\centerline{\includegraphics[width=.95\columnwidth]{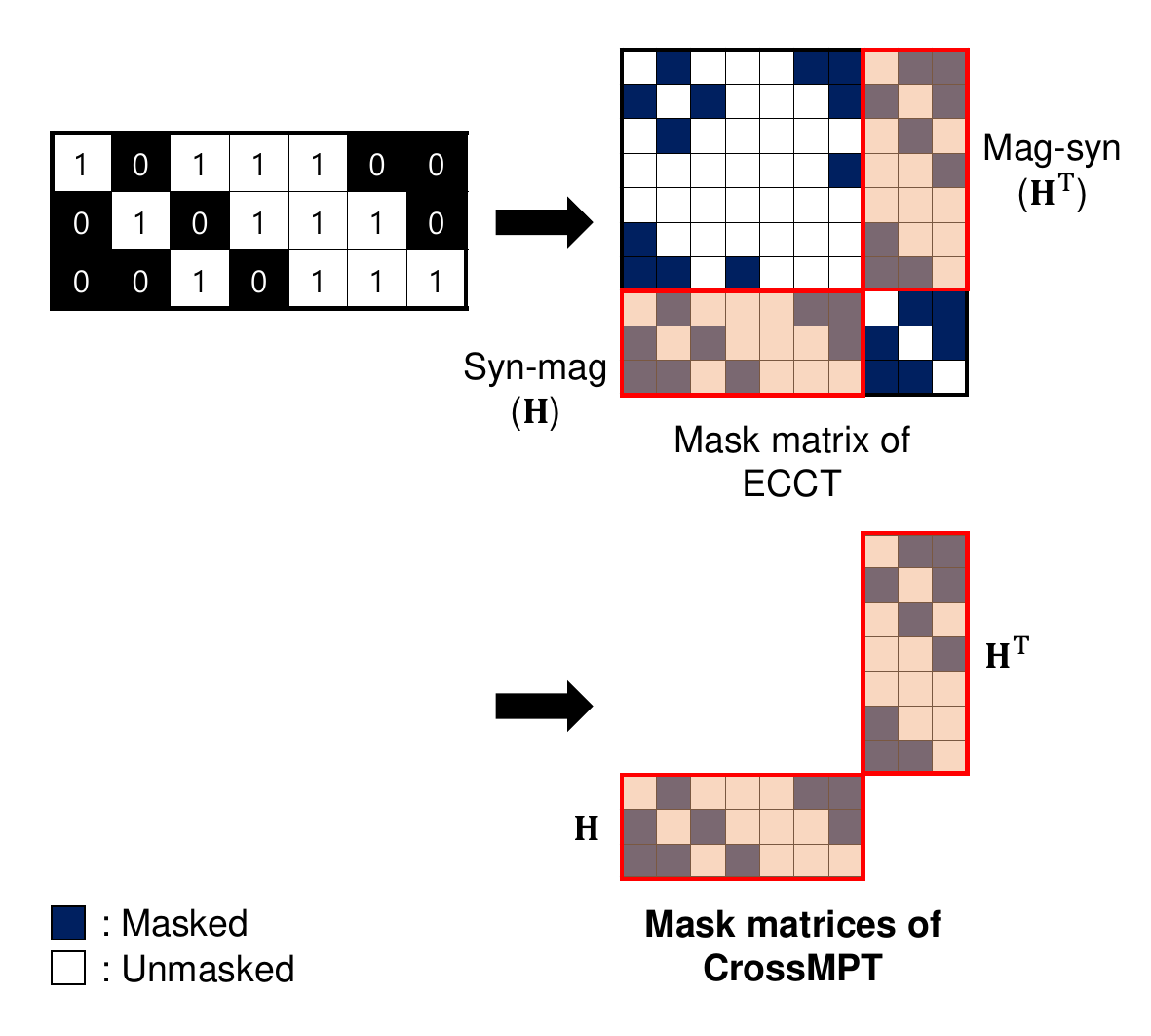}}
\caption{The PCMs and the mask matrices of ECCT~\cite{b_ECCT} and CrossMPT.\label{fig_mask}}
\end{center}
\end{figure}

\subsection{Error Correction Code Transformer}

ECCT is the first approach to present a model-free decoder with the transformer architecture.
ECCT outperforms other neural decoders by employing the masked self-attention mechanism, whose mask matrix is determined by the code's PCM~\cite{b_ECCT}.
A primary challenge in training transformer-based decoders is the issue of overfitting.
In \cite{b_Bennatan2018}, the overfitting issue in model-free neural decoders is described as poor generalization to untrained codewords due to the exponentially large number of possible codewords.
However, it has been resolved by a preprocessing technique that facilitates syndrome-based decoding~\cite{b_Bennatan2018}. 
It has been theoretically proven that, with this preprocessing step, the decoder’s performance remains invariant to the specific codewords used in the training set~\cite{b_Bennatan2018}.

As in~\cite{b_Bennatan2018}, the preprocessing step of ECCT utilizes the magnitude and syndrome vectors to learn multiplicative noise $\tilde{\mathbf{z}}_s$, which is defined by
\begin{equation}
\label{equ_multi_noise}
    \mathbf{y}=\mathbf{x}_s+\mathbf{z} = \mathbf{x}_s\tilde{\mathbf{z}}_s.
\end{equation}

\begin{figure*}[!t]
\begin{center}
\subfloat[]{\includegraphics[width=.36\textwidth]{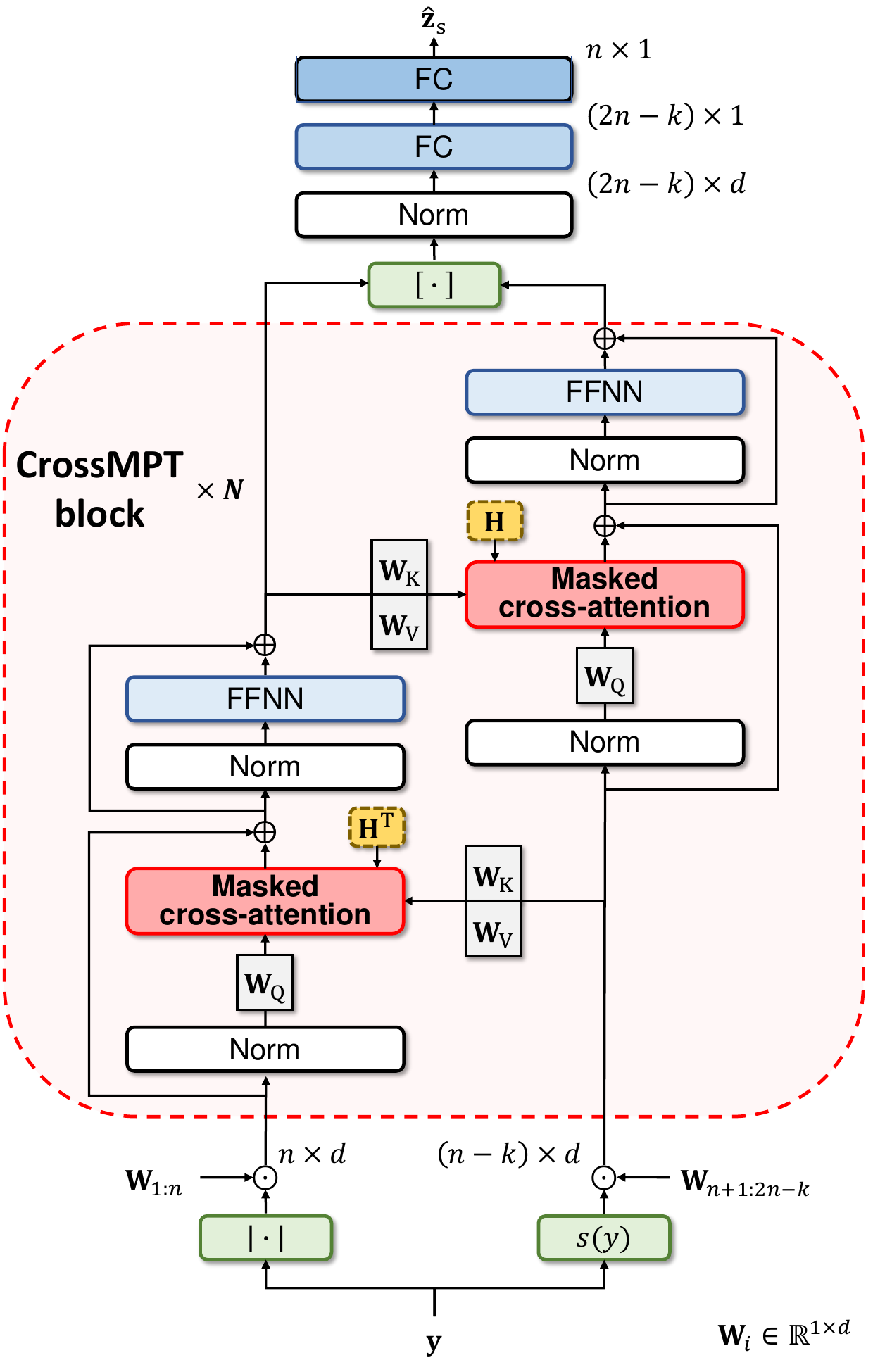}}
\hspace{0.12cm} 
\hfil
\subfloat[]{\includegraphics[width=.36\textwidth]{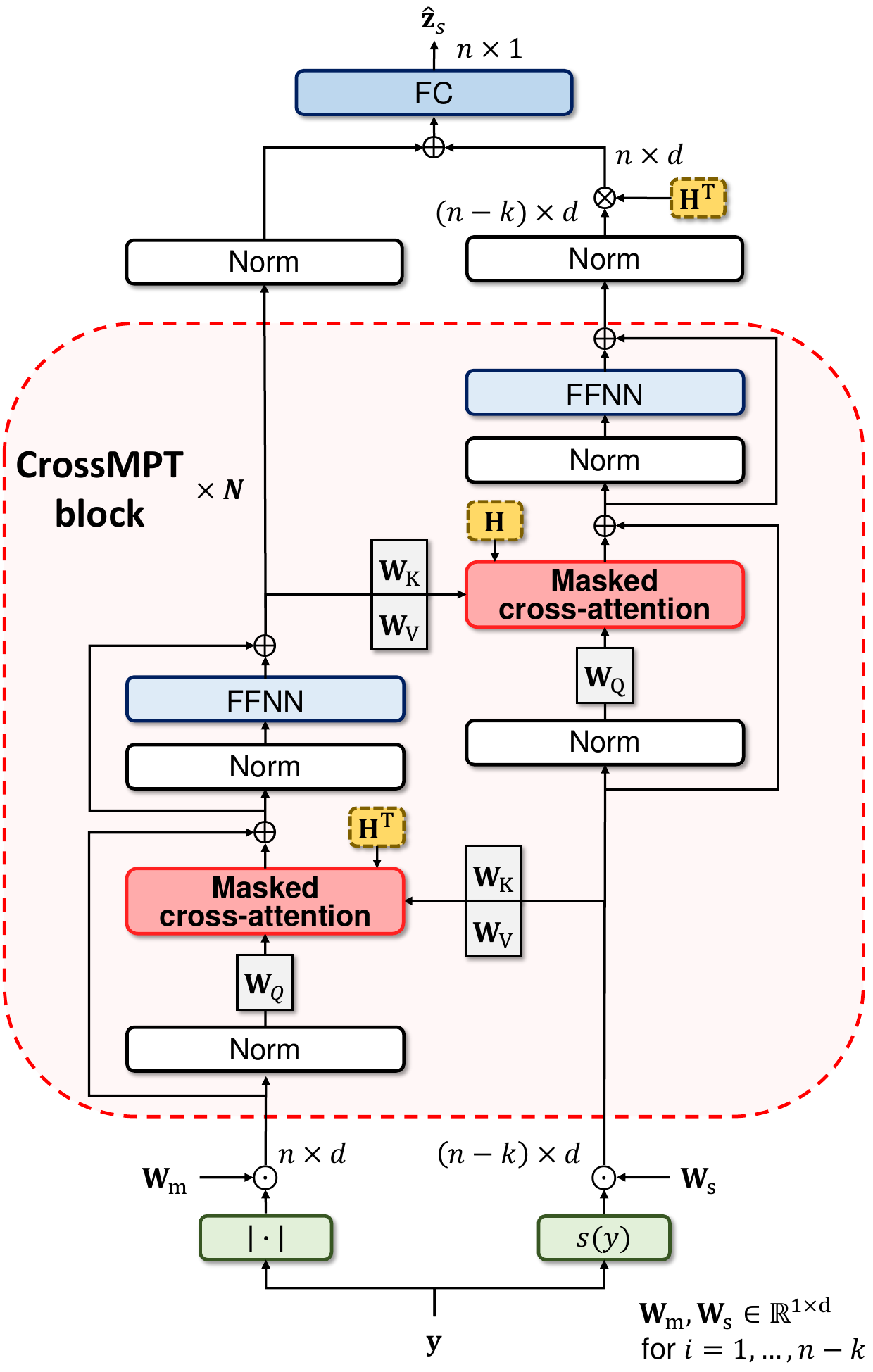}}
\caption{The architectures of (a) CrossMPT and (b) foundation CrossMPT (FCrossMPT).
\label{fig_archi}}
\end{center}
\vspace{-4mm}
\end{figure*}

ECCT aims to estimate the multiplicative noise in (\ref{equ_multi_noise}), i.e.,  \mbox{$f(\mathbf{y})=\hat{\mathbf{z}}_s$}.
Then, the estimation of $x$ is
\mbox{$\hat{\mathbf{x}} = \text{bin}(\text{sign}(\mathbf{y}f(\mathbf{y})))$}.
If the multiplicative noise is correctly estimated such that \mbox{$\text{sign}(\tilde{\mathbf{z}}_s)=\text{sign}(\hat{\mathbf{z}}_s)$}, then $\hat{\mathbf{x}}$ can be computed as:
\begin{equation*}
    \hat{\mathbf{x}} = \text{bin}(\text{sign}(\mathbf{y}f(\mathbf{y})))= \text{bin}(\text{sign}(\mathbf{x}_s\tilde{\mathbf{z}}_s\hat{\mathbf{z}}_s))=\text{bin}(\text{sign}(\mathbf{x}_s)).
\end{equation*}

ECCT employs a masked self-attention module to train the transformer architecture, where the input embedding is formed by concatenating the magnitude and syndrome embeddings.
As shown in Fig.~\ref{fig_mask}, the mask matrices of ECCT are designed to clearly distinguish between necessary (unmasked) and unnecessary (masked) pairwise relationships among magnitude-magnitude, magnitude-syndrome, and syndrome-syndrome bit pairs.
In ECCT, the syndrome-syndrome part is unmasked only for self-relations, while the magnitude-syndrome part is unmasked based on the connections derived from the PCM.
The magnitude-magnitude part, however, is unmasked for bit pairs that are connected at depth~2 (see Algorithm 1 in \cite{b_ECCT}).
While the masking of magnitude-syndrome relations is straightforward, as it directly uses PCM, identifying meaningful relations among magnitude bits is not directly derivable from the PCM.
In Fig.~\ref{fig_mask}, the white areas indicate unmasked positions that require attention calculations, while the blue areas represent masked positions where attention calculations can be omitted. 
As the proportion of masked (blue) entries increases, the attention matrix becomes sparser, leading to improved computational efficiency.

\section{Proposed Methods}
\label{sec_proposed}
In this section, we present the architecture of CrossMPT with its operational mechanism.
CrossMPT processes the magnitude and syndrome embeddings separately and applies a cross-attention mechanism to effectively capture their distinct information.
It shares core principles with message passing algorithms for decoding linear codes by iteratively updating the magnitude and syndrome embeddings of the received vector.
The overall architecture is depicted in Fig.~\ref{fig_archi}(a).

\begin{figure}[!t]
\begin{center}
\centerline{\includegraphics[width=\columnwidth]{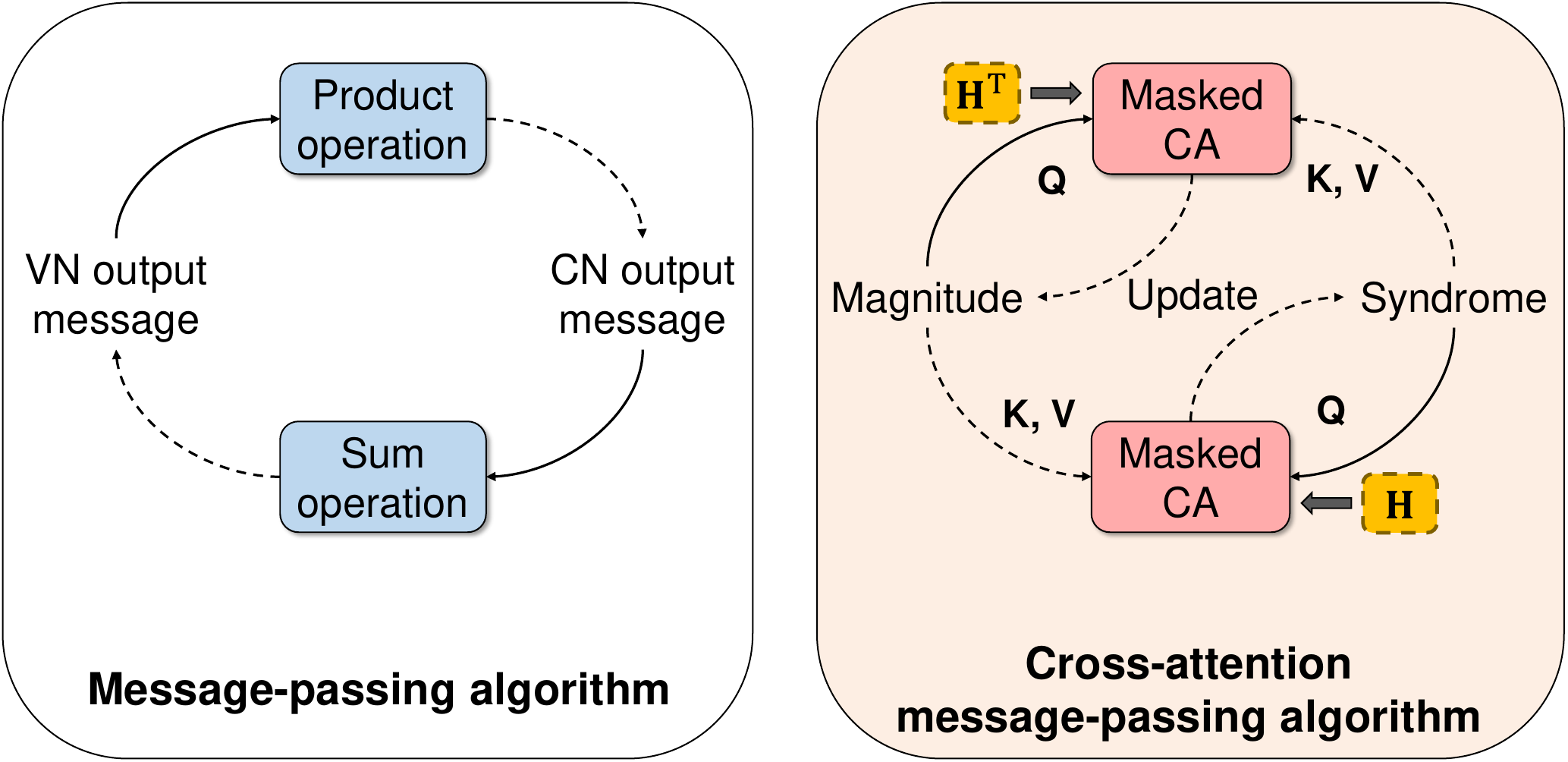}}
\caption{Conceptual comparison of the sum-product message-passing algorithm and the proposed cross-attention~(CA) message-passing algorithm.\label{fig_message_passing}}
\end{center}
\end{figure}

\subsection{Cross-Attention Message-Passing Transformer}


The CrossMPT architecture has two cross-attention blocks.
One cross-attention block updates the magnitude embedding by using it as the query while the syndrome embedding serves as the key and value.
This results in an attention map of size $n\times (n-k)$, representing the `magnitude-syndrome' relationships.
We employ the transpose of the PCM $\mathbf{H}^{\top}$ as the mask matrix.
This is because the $n$ rows of $\mathbf{H}^{\top}$ correspond to the $n$ bit positions, and its $n-k$ columns correspond to the parity check equations, directly aligning with $|\mathbf{y}|$ and $s(\mathbf{y})$, respectively.
In the other cross-attention block, the syndrome embedding is used as the query, while the magnitude embedding provides the key and value.
For this block, we utilize the PCM $\mathbf{H}$ as the mask matrix.

This design—separately processing two distinct informational components—mirrors the structure of message-passing algorithms used in decoding linear codes.
Message-passing algorithms such as the sum-product algorithm~\cite{b_BP} are widely adopted for decoding LDPC codes.
The sum-product algorithm operates by exchanging messages between variable nodes~(VNs) and check nodes~(CNs) over a Tanner~(bipartite) graph~\cite{b_BP}.
In the Tanner graph, VNs represent bits of the received vector and convey information about bit reliability, while CNs correspond to the parity check equations.
The edges between VNs and CNs represent the connections~(relationships) between them.
The sum-product algorithm iteratively updates the output messages of VNs and CNs by exchanging messages along these edges.

Similar to this principle, CrossMPT performs iterative updates of the magnitude and syndrome embeddings.
First, the magnitude embedding is updated using a masked cross-attention block, where the magnitude embedding serves as the query input and syndrome embedding provides the key and value inputs.
Next, the updated magnitude embedding is used to update the syndrome embedding in a subsequent masked cross-attention block, where the syndrome embedding serves as the query input and the updated magnitude embedding provides the key and value inputs.
This iterative process enables CrossMPT to progressively refine both embeddings, resulting in more accurate estimation of the multiplicative noise.

Fig.~\ref{fig_message_passing} compares the sum-product and the proposed cross-attention message-passing algorithms.
In the sum-product algorithm, the VN output and CN output messages are updated through summation and multiplication operations.
Similarly, in CrossMPT, the magnitude and syndrome embeddings are iteratively updated using the masked cross-attention~(denoted as `Masked CA' in the figure) blocks in CrossMPT.
Note that $\mathbf{H}$ and $\mathbf{H}^\top$ are used as the mask matrices in the cross-attention blocks, and $\mathbf{Q}$, $\mathbf{K}$, and $\mathbf{V}$ denote the query, key, and value in the cross-attention mechanism.

\subsection{Model Architecture of CrossMPT}

In the initial embedding layer, we compute $|\mathbf{y}|=(|y_1|,\ldots,|y_n|)$ and $s(\mathbf{y})=(s(\mathbf{y})_1,\ldots ,s(\mathbf{y})_{n-k})$ from the received vector $\mathbf{y}$. The elements $y_i$ and $s(\mathbf{y})_i$ are then projected into $d$ dimensional embedding row vectors $\mathbf{M}_i$ and $\mathbf{S}_i$, respectively, as follows:
\begin{align*}
    \mathbf{M}_i&=|y_i|\mathbf{W}_i,\qquad~~\text{ for } i=1,\ldots,n,\\
    \mathbf{S}_i&=s(\mathbf{y})_i \mathbf{W}_{i+n},\quad \text{ for } i=1,\ldots,n-k,
\end{align*}
where $\mathbf{W}_i \in \mathbb{R}^{1\times d}$ for $i=1,\ldots,2n-k$ denotes the trainable positional encoding vector.

These magnitude and syndrome embeddings are processed as separate inputs in the subsequent $N$ decoding layers.
Each decoding layer contains two cross-attention blocks, each consisting of a cross-attention module, a feed-forward neural network (FFNN), and a normalization layer.

In the first cross-attention module, the magnitude embedding is updated using the syndrome embedding.
The query $\mathbf{Q}_1$, key $\mathbf{K}_1$, and value $\mathbf{V}_1$ are assigned as follows:
\begin{equation*}
\mathbf{Q}_1=\mathbf{M}\mathbf{W}_{\rm Q}, \mathbf{K}_1=\mathbf{S}\mathbf{W}_{\rm K}, \mathbf{V}_1=S\mathbf{W}_{\rm V},
\end{equation*}
where $\mathbf{M}=[ \mathbf{M}_1; \cdots; \mathbf{M}_n]\in\mathbb{R}^{n\times d}$ and $\mathbf{S}=[\mathbf{S}_1; \cdots ; \mathbf{S}_{n-k}]\in\mathbb{R}^{(n-k)\times d}$ denote the magnitude and syndrome embeddings, respectively. 
Here, $\mathbf{W}_{\rm Q}$, $\mathbf{W}_{\rm K}$, and $\mathbf{W}_{\rm V}$ are the learnable projections for the query, key, and value, respectively.
This operation is classified as \emph{cross-attention} since the query contains different information from the key and value.
The scaled dot-product attention is then computed as:
\begin{equation*}
{\rm Attention}(\mathbf{Q}_1,\mathbf{K}_1,\mathbf{V}_1)={\rm softmax}\left(\dfrac{\mathbf{Q}_1\mathbf{K}_1^{\top}+g(\mathbf{H}^{\top})}{\sqrt{d}}\right)\mathbf{V}_1,
\end{equation*}
where $g(\mathbf{H}^{\top})$ is the mask matrix, and the function $g$ is defined as
\begin{equation}
    g(\mathbf{A})_{i,j}=\begin{cases} 0 \qquad~{\rm if }~\mathbf{A}_{i,j}=1,\\ -\infty\quad{\rm if }~\mathbf{A}_{i,j}=0. \end{cases}
\end{equation}

This configuration results in an attention map of size $n\times(n-k)$, which captures the `magnitude-syndrome' relationship.
Therefore, we use the transpose of the PCM $\mathbf{H}^{\top}$ as the mask matrix.
The output of this cross-attention module is the updated magnitude embedding $\mathbf{M}'$.

In the second cross-attention block, we update the `syndrome' embedding using the updated magnitude embedding $\mathbf{M}'$. 
In other words, the syndrome embedding serves as the query input, while $\mathbf{M}'$ is used for both the key and value inputs. 
We reuse the \textit{same projections} $\mathbf{W}_{\rm Q}$, $\mathbf{W}_{\rm K}$, and $\mathbf{W}_{\rm V}$ from the first cross-attention module. 
The query $\mathbf{Q}_2$, key $\mathbf{K}_2$, and value $\mathbf{V}_2$ are defined as follows:
\begin{equation*}
    \mathbf{Q}_2=\mathbf{S}\mathbf{W}_{\rm Q}, \mathbf{K}_2=\mathbf{M}'\mathbf{W}_{\rm K}, \mathbf{V}_2=\mathbf{M}'\mathbf{W}_{\rm V}.
\end{equation*}
Since the syndrome and magnitude embeddings correspond to the rows and columns of the attention map, respectively, we use the mask matrix $g(\mathbf{H})$, where the zero entries in $\mathbf{H}$ define the masked positions.
Then, we apply the scaled dot-product attention and the resulting output is the updated syndrome embedding $\mathbf{S}'$.
This updated embedding is subsequently used in the next decoder layer to further refine the magnitude embedding.
This iterative process continues across all $N$ decoder layers.

Finally, the updated magnitude and syndrome embeddings from the last decoder layer are concatenated and passed through a normalization layer and two fully connected~(FC) layers.
The first FC layer reduces the $(2n-k) \times d$ dimensional embedding to a $(2n-k)$-dimensional vector, and the second FC layer further maps this vector to a vector of length $n$. 
The final output provides the estimation of $\tilde{\mathbf{z}}_s$.
Since the two cross-attention blocks of CrossMPT share the same weight matrices $\mathbf{W}_{\rm Q}, \mathbf{W}_{\rm K}, \mathbf{W}_{\rm V}$, and all other layers are identical to those in ECCT, CrossMPT retains the same number of parameters as the original ECCT.

\begin{figure}[!t]
\begin{center}
\centerline{\includegraphics[width=.65\columnwidth]{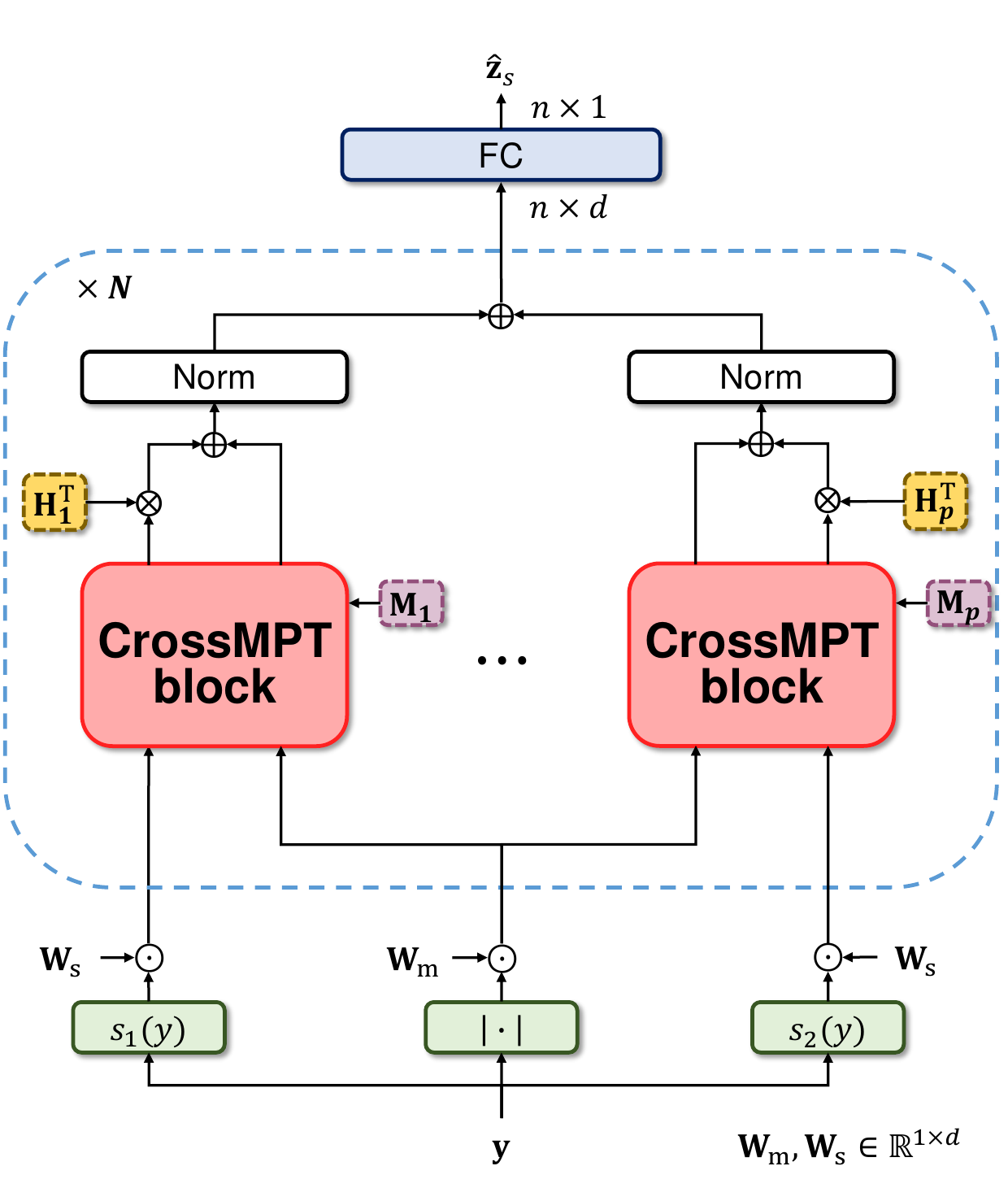}}
\caption{The architecture of CrossED.\label{fig_CrossED}}
\end{center}
\vspace{-4mm}
\end{figure}

\begin{figure}[!t]
\begin{center}
\centerline{\includegraphics[width=.5\columnwidth]{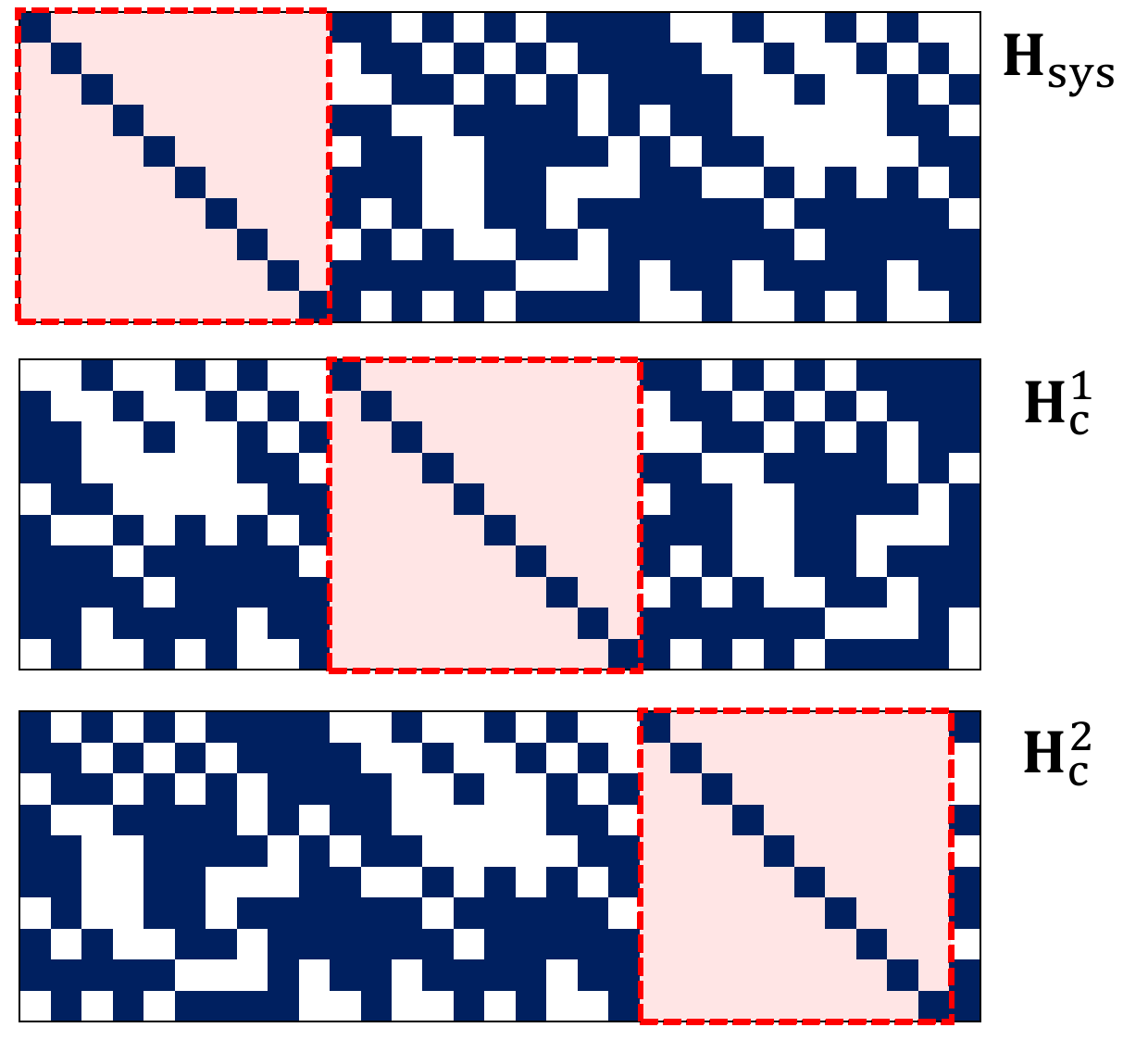}}
\caption{Three PCMs of $(31,21)$ BCH codes utilized in $p=3$ CrossED.\label{fig_TM_PCM}}
\end{center}
\vspace{-6mm}
\end{figure}

\subsection{Foundation CrossMPT}

We propose a foundation model of CrossMPT, referred to as FCrossMPT, by replacing the code-dependent parameters of CrossMPT with code-invariant embeddings.
FCrossMPT retains the core architecture of CrossMPT, including its cross-attention blocks and iterative message-passing structure.
However, it introduces two key modifications to ensure invariance to code length and rate, thereby enabling generalization across a wide range of codes.

First, to make the architecture code-agnostic, we adopt the principle from FECCT~\cite{b_Choukroun2024} for the initial embedding layer.
Instead of training position-specific embeddings for each magnitude and syndrome element, FCrossMPT employs a single shared embedding vector for all magnitude elements and another for all syndrome elements.
While FECCT applies this concept to a self-attention architecture, our FCrossMPT integrates this foundation model principle into our novel cross-attention message-passing framework, thereby preserving its computational advantages.
Each input vector $|\mathbf{y}| = (|y_1|,\ldots,|y_n|)$ and $s(\mathbf{y}) = (s(\mathbf{y})_1,\ldots,s(\mathbf{y})_{n-k})$ is embedded as follows:
\begin{align*}
    \mathbf{M}_i&=|y_i|\mathbf{W}_{\rm M},\qquad ~\text{ for } i=1,\ldots,n,\\
    \mathbf{S}_i&=s(\mathbf{y})_i\mathbf{W}_{\rm S},\qquad \text{ for } i=1,\ldots,n-k,
\end{align*}
where $\mathbf{W}_{\rm M}, \mathbf{W}_{\rm S} \in \mathbb{R}^{1 \times d}$ are the shared trainable embedding vectors, independent of positional index. 
This design ensures invariance to both the code length $n$ and dimension $k$, thereby enabling unified decoding across various codes.

Second, in the output layer, FCrossMPT again replaces code-specific parameters with code-invariant ones.
The two output embeddings~(magnitude and syndrome embeddings) from the last decoder layer are passed through a normalization layer.
The syndrome embedding, originally of shape $(n-k) \times d$, is resized to match the magnitude embedding of shape $n \times d$ by multiplying it with the PCM $\mathbf{H}^{\top}$.
The resized syndrome embedding is then added to the magnitude embedding.
This approach is motivated by findings from FECCT~\cite{b_Choukroun2024}, which shows that each prediction bit can be expressed as a linear combination of magnitude and syndrome embeddings.

Through these modifications, FCrossMPT ensures that all model parameters are invariant to position and length, allowing the model to function as a foundation decoder capable of decoding various codes without architecture changes or retraining.

\subsection{CrossMPT Ensemble Decoder}

To additionally enhance the decoding performance, we propose an ensemble decoding architecture called CrossMPT ensemble decoder, or CrossED, as shown in Fig.~\ref{fig_CrossED}.
The proposed CrossED consists of $p$ parallel CrossMPT blocks, each employing a different mask matrix.
The output of these blocks are fused with addition operation in the output layer.

Let $\mathbf{H}_j$ and $\mathbf{M}_j$ denote the $j$th PCM and mask matrix, respectively, used in CrossED.
During the initial embedding stage, we first generate the magnitude vector $|\mathbf{y}|=(|y_1|,{\ldots},|y_n|)$ and $p$ distinct syndrome vectors $s_j(\mathbf{y})=(s_j(\mathbf{y})_1,{\ldots},s_j(\mathbf{y})_{n-k})$, where $s_j(\mathbf{y})=\mathbf{H}_j\mathbf{y}_b$ for $j=1,\ldots,p$.
Then, each element of $|\mathbf{y}|$ and $s_j(\mathbf{y})$'s is projected into $d$ dimension embedding vectors $\mathbf{M}_i$ and $\mathbf{S}_{j,i}$, respectively, as follows:
\begin{align}
\label{eq_CrossED}
    \mathbf{M}_i&=|y_i|\mathbf{W}_{\rm M},\;\;\;\text{ for } i=1,\ldots,n,\\
    \mathbf{S}_{j,i}&=s_j(\mathbf{y})_i\mathbf{W}_{\rm S}, \text{ for } i=1,\ldots,n-k\text{ and }j=1,\ldots,p,
\end{align}
where $\mathbf{W}_{\rm M},\mathbf{W}_{\rm S} \in \mathbb{R}^{1\times d}$ denote the trainable positional encoding vectors for magnitude embedding and syndrome embedding, respectively.
As shown in (\ref{eq_CrossED}), all $s_j(\mathbf{y})$'s share the same embedding weight $\mathbf{W}_{\rm S}$.
Each decoder layer consists of $p$ parallel CrossMPT blocks, where each block utilizes its own distinct mask matrix.
The $p$ parallel CrossMPT blocks are repeated over $N$ decoder layers.

In the output layer, each CrossMPT block produces two output embeddings (magnitude and syndrome), which are normalized and fused.
The syndrome embedding $S_{j}$ is resized from $(n-k) \times d$ to $n \times d$ via multiplication with the corresponding PCM $\mathbf{H}_j$ for $j=1,\ldots,p$, as in FCrossMPT.
Finally, all $p$ different $n \times d$ embeddings are added and pass through a FC layer, yielding an estimate of the multiplicative noise with dimension $n \times 1$.

In CrossED, each CrossMPT block employs the systematic PCM and its complementary PCMs as a mask matrix, which are first introduced in MM-ECCT~\cite{Park2024multiple}.
The systematic PCM $\mathbf{H}_{\rm sys}$ is expressed as $\mathbf{H}_{\rm sys} = [\mathbf{I}_{n-k}~\mathbf{P}]$, where $\mathbf{I}_{n-k}$ is the identity matrix of size $(n-k)\times(n-k)$ and $\mathbf{P}$ is a matrix of size $(n-k)\times k$.
The complementary PCM of $\mathbf{H}_{\rm sys}$ includes an identity matrix of size $(n-k) \times (n-k)$ spanning from bit positions $n-k+1$ to $2n-k$.
A complementary PCM is then constructed to have its identity matrix in a different position to the original $\mathbf{H}_{\rm sys}$, allowing the ensemble to collectively cover weaknesses across all bit positions.
An example of a systematic PCM and its complementary versions for a $(31,21)$ BCH code is shown in Fig.~\ref{fig_TM_PCM}.

For non-cyclic codes, if a full identity matrix cannot be formed in the desired positions through row operations, we aim to diagonalize that portion of the matrix as much as possible using methods like Gaussian elimination.
For cyclic codes, however, this process is significantly simplified. Based on the property that a cyclic shift of a codeword remains a valid codeword, the complementary PCM $\mathbf{H}^p_c(i,j)$ can be readily obtained by cyclically shifting the columns of the systematic PCM, as defined in (\ref{equ_H_c}).
\begin{equation}
\label{equ_H_c}
    \mathbf{H}^p_c(i,j) = \mathbf{H}_{\rm sys}(i,(j-p(n-k))\text{ mod }n),
\end{equation}
where $p=1,{\ldots},\lceil n/(n-k) \rceil-1$, $0\le i < n-k$, and $0\le j < n$.

\begin{figure*}[!t]
\begin{center}
\subfloat[]{\includegraphics[width=.45\textwidth]{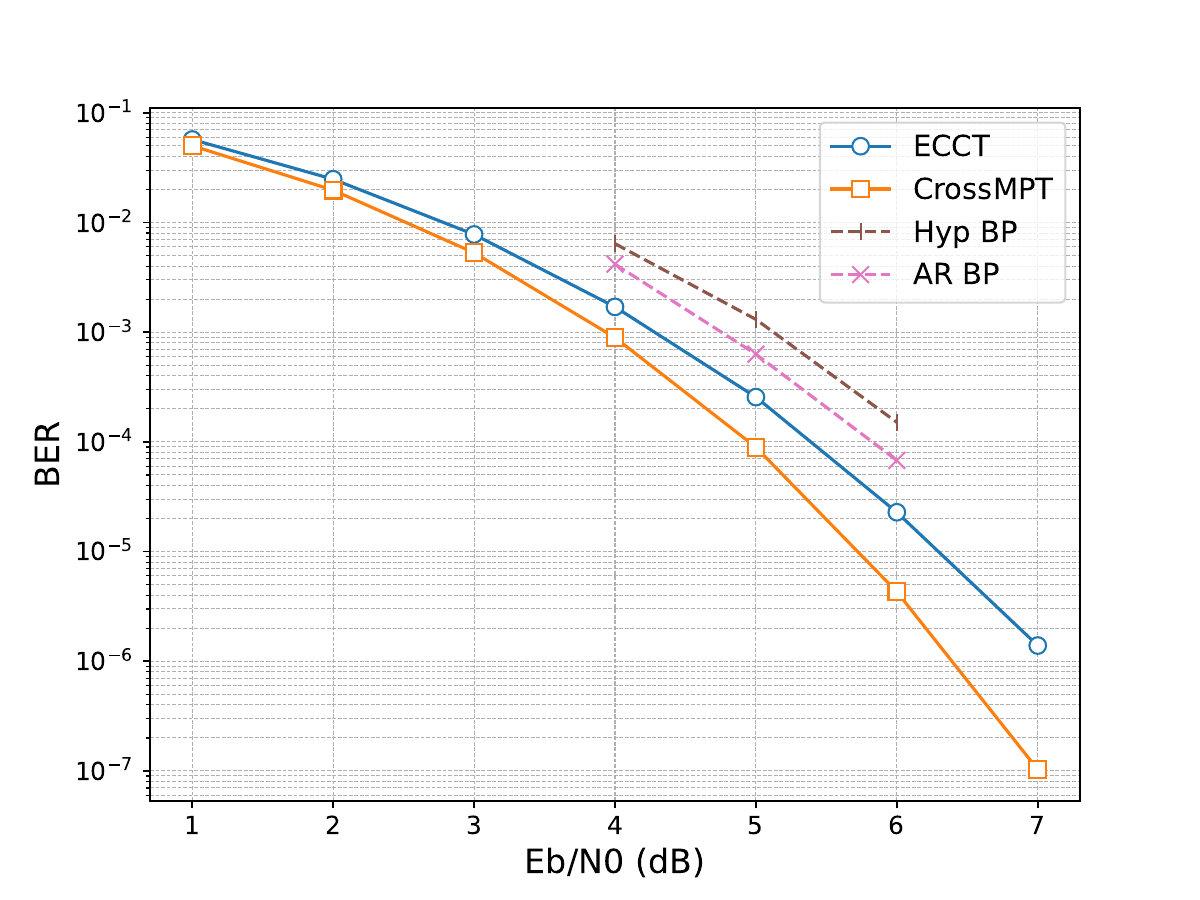}}
\subfloat[]{\includegraphics[width=.45\textwidth]{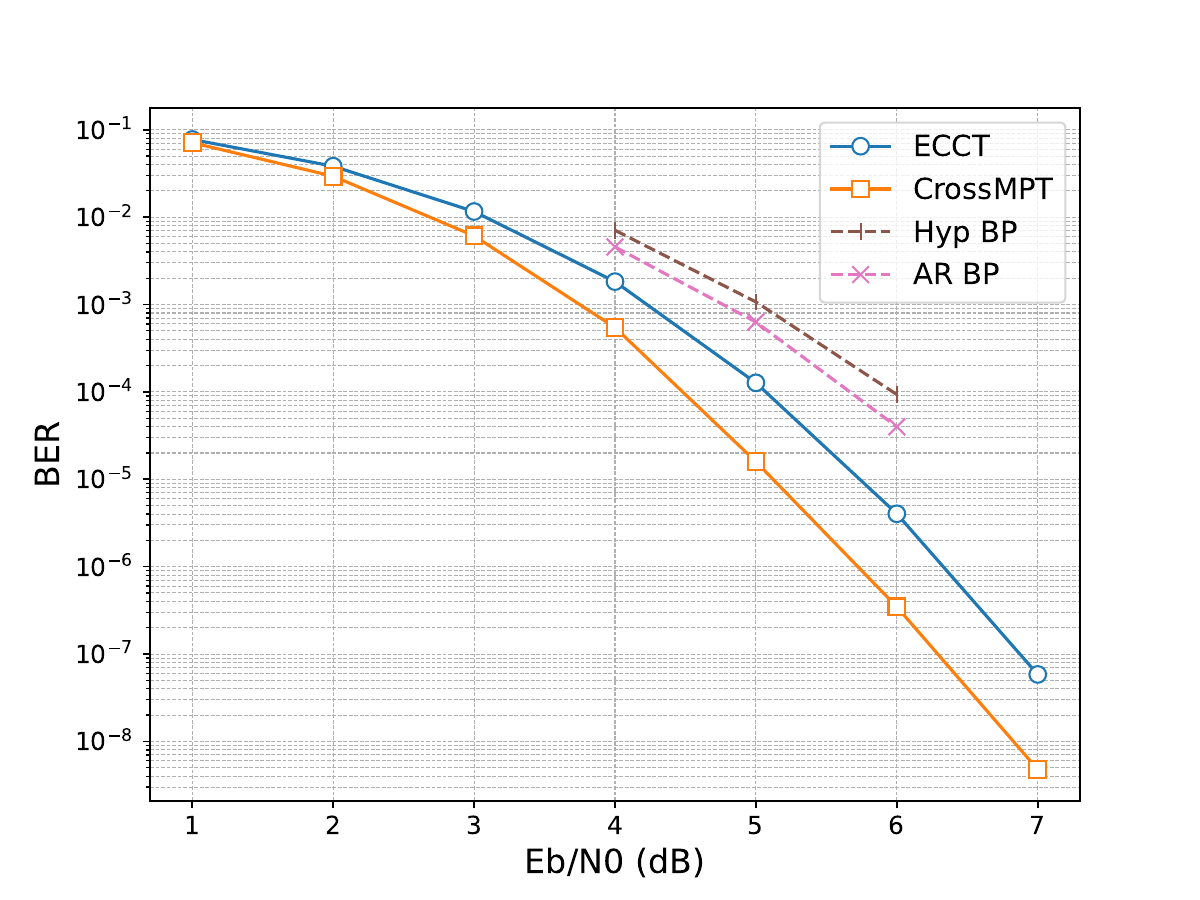}}
\caption{BER performances of ECCT, CrossMPT, and BP-based neural decoders for  (a) $(31, 16)$ BCH code and (b) $(128, 86)$ polar code.
\label{fig_graph}}
\end{center}
\vspace{-4mm}
\end{figure*}

\begin{figure*}[!t]
\begin{center}
\subfloat[\label{fig_larger_graph_a}]{\includegraphics[width=.45\textwidth]{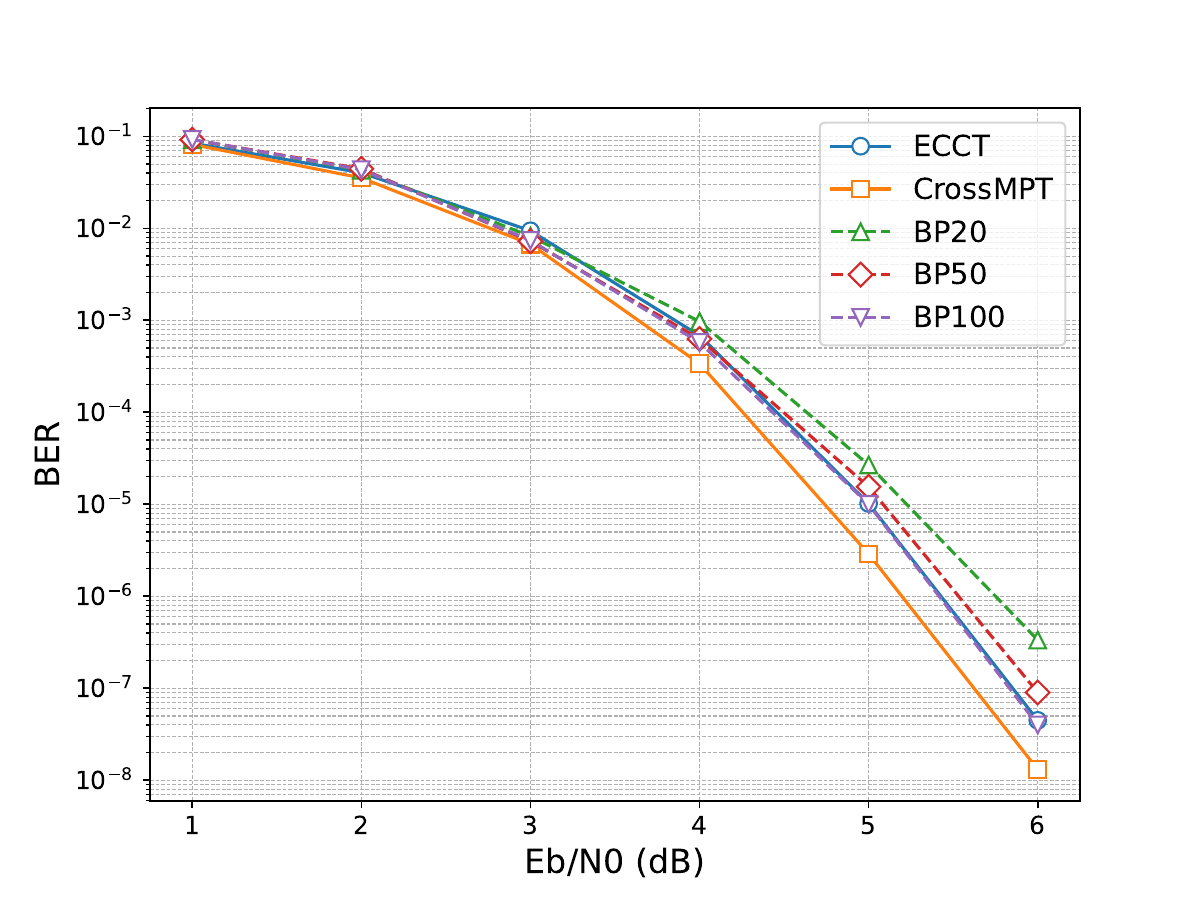}}
\subfloat[\label{fig_larger_graph_b}]{\includegraphics[width=.45\textwidth]{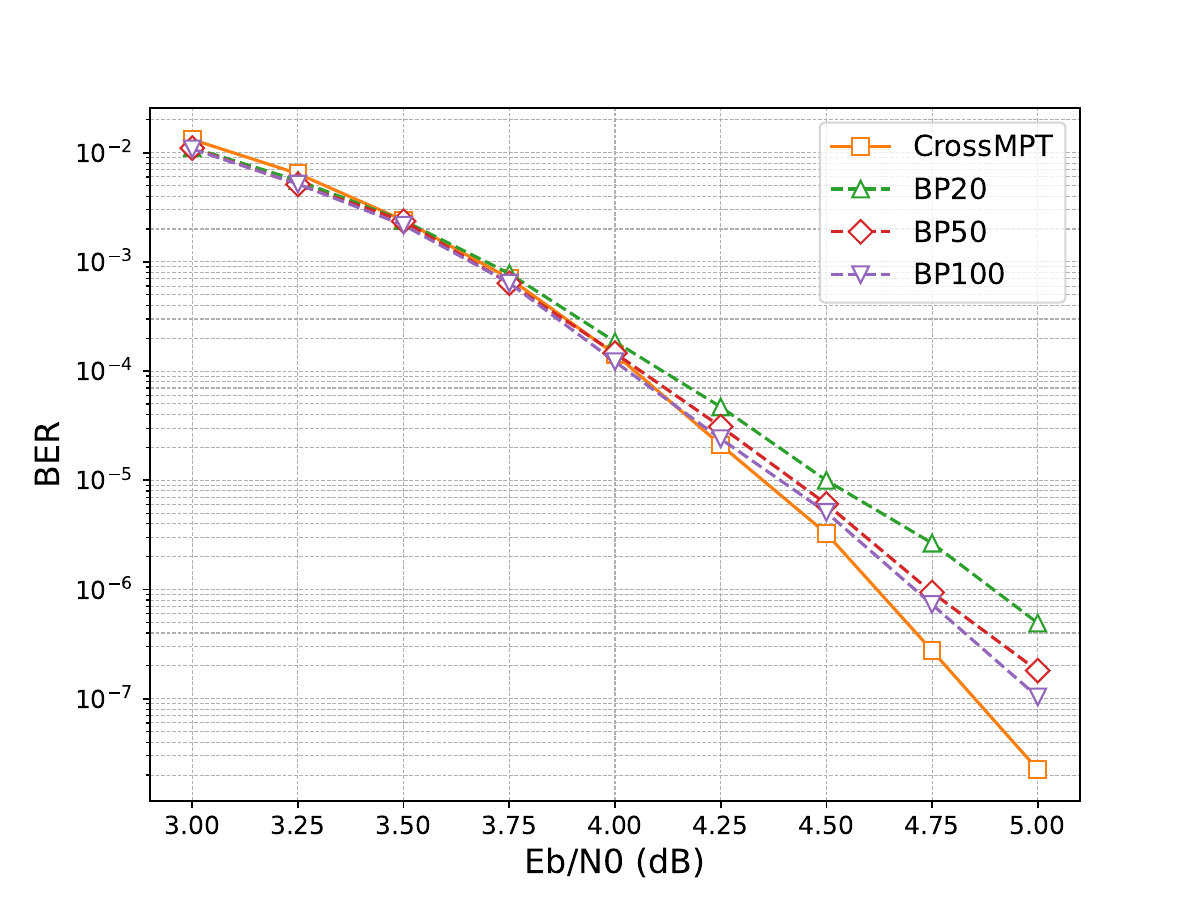}}
\caption{BER performances of BP decoders~(iteration 20, 50, and 100) and CrossMPT for  (a) $(121,80)$ LDPC code and (b) $(648,540)$ IEEE 802.11n LDPC code.
\label{fig_larger_BP}}
\end{center}
\end{figure*}

It is shown that the bit error rate~(BER) of positions corresponding to the identity matrix $\mathbf{I}_{n-k}$ of the PCM tends to be lower than that of other positions~\cite{Park2024multiple}.
Thus, when the systematic PCM and complementary PCMs, whose identity matrix positions differ, are employed together in CrossED, they supplement each other and boost the decoding performance.
When the systematic and complementary PCMs are utilized in CrossED, the first $n-k$ bits are covered by the identity matrix of the systematic PCM, while the subsequent $n-k$ bits are covered by the identity matrix of the complementary PCM.
To ensure that the identity parts cover all $n$ bits, a total of $\lceil n/(n-k) \rceil$ PCMs are required.
CrossED also can be served as a foundation decoder for ECCs since all parameters in the decoding process are position-invariant and length-invariant.

\subsection{Training}
\label{sec_training}

The goal of the proposed decoder is to learn the multiplicative noise $\tilde{\mathbf{z}}_s$ in (\ref{equ_multi_noise}) and estimate the transmitted signal $\mathbf{x}_s$.
The multiplicative noise can be computed as $\tilde{\mathbf{z}}_s = \mathbf{y}\mathbf{x}_s = \tilde{\mathbf{z}}_s \mathbf{x}^2_s$.
Accordingly, the target for the binary cross-entropy loss is defined as $\tilde{\mathbf{z}} = \text{bin}(\text{sign}(\mathbf{y}\mathbf{x}_s))$.
The corresponding loss function for a received vector $\mathbf{y}$ is given by:
\begin{equation*}
    \mathcal{L}=-\sum_{i=1}^n \{\tilde{ \mathbf{z}}_i\log(1-\sigma(f(\mathbf{y})_i))+(1-\tilde{ \mathbf{z}}_i)\log( \sigma(f(\mathbf{y})_i))\}.
\end{equation*}

For a fair comparison, we adopt the same training setup used in the previous work~\cite{b_ECCT}.
We use the Adam optimizer~\cite{b_adam} and conduct 1000 epochs.
Each epoch consists of 1000 minibatches, where each minibatch is composed of 128 samples.
All simulations were conducted using NVIDIA GeForce RTX 3090 GPU and AMD Ryzen 9 5950X 16-Core Processor CPU.
The training sample $\mathbf{y}$ is generated by $y=\mathbf{x}_s+\mathbf{z}$, where $\mathbf{x}_s$ is the modulated signal corresponding to the all-zero codeword,  and $\mathbf{z}$ represents the AWGN channel noise, sampled from an normalized signal-to-noise ratio~($\rm E_b/N_0$) range of $3$~dB to $7$~dB.
The learning rate is initially set to $10^{-4}$ and gradually reduced to $5\times10^{-7}$ following a cosine decay scheduler.

\section{Experimental Results}

In this section, we compare experimental results for four different proposed decoders: CrossMPT, FCrossMPT, CrossED, and foundation CrossED~(FCrossED).
Our results include a comparison with conventional transformer-based decoders trained with a single type of code class.

\subsection{Decoding Performance of Code-Specific Decoders}

We compare the BER performance of CrossMPT with the BP-based neural decoders~\cite{b_Nachmani2019,b_Nachmani2021}, and ECCT~\cite{b_ECCT} in Fig.~\ref{fig_graph}.
All PCMs are taken from~\cite{b_channelcode}.
The results for both for the (31, 16) BCH code and  (128, 86) polar code. CrossMPT and ECCT are obtained with $N=6$ and $d=128$.
As shown in the figure, CrossMPT outperforms the conventional ECCT and all the other BP-based neural decoders.

Moreover, Fig.~\ref{fig_larger_BP} shows the performance comparison between the traditional BP decoder with a maximum number of iterations of 20, 50, and 100 and CrossMPT for the $(121,80)$ short LDPC code and $(648,540)$ IEEE 802.11n LDPC code.
We configure ECCT and CrossMPT with $N=6$ and $d=128$ for the $(121,80)$ LDPC code, and with $N=10$ and $d=128$ for the $(648,540)$ IEEE 802.11n LDPC code. We do not include a comparison with ECCT for the long code in Fig.~\ref{fig_larger_BP}(b), as the ECCT model could not be implemented in our simulation environment due to its high memory consumption.
Notably, the proposed CrossMPT outperforms the conventional BP decoder for both short and long LDPC codes.
Considering that the BP decoder is specifically designed for LDPC codes and leverages well-established iterative algorithms, it is remarkable that CrossMPT, trained solely through data-driven learning without any domain-specific algorithmic prior, achieves superior performance.

An important aspect of our research is CrossMPT's capability to decode long codes as shown in Fig.~\ref{fig_larger_BP}(b), which remain beyond the reach of ECCT due to its high memory requirements, resulting from large attention maps.
These results demonstrate the practical significance and architectural advantages of CrossMPT, proving its value in scenarios where ECCT encounters limitations.

\begin{figure*}[!t]
\begin{center}
\subfloat[\label{fig_121_70_FCross}]{\includegraphics[width=.45\textwidth]{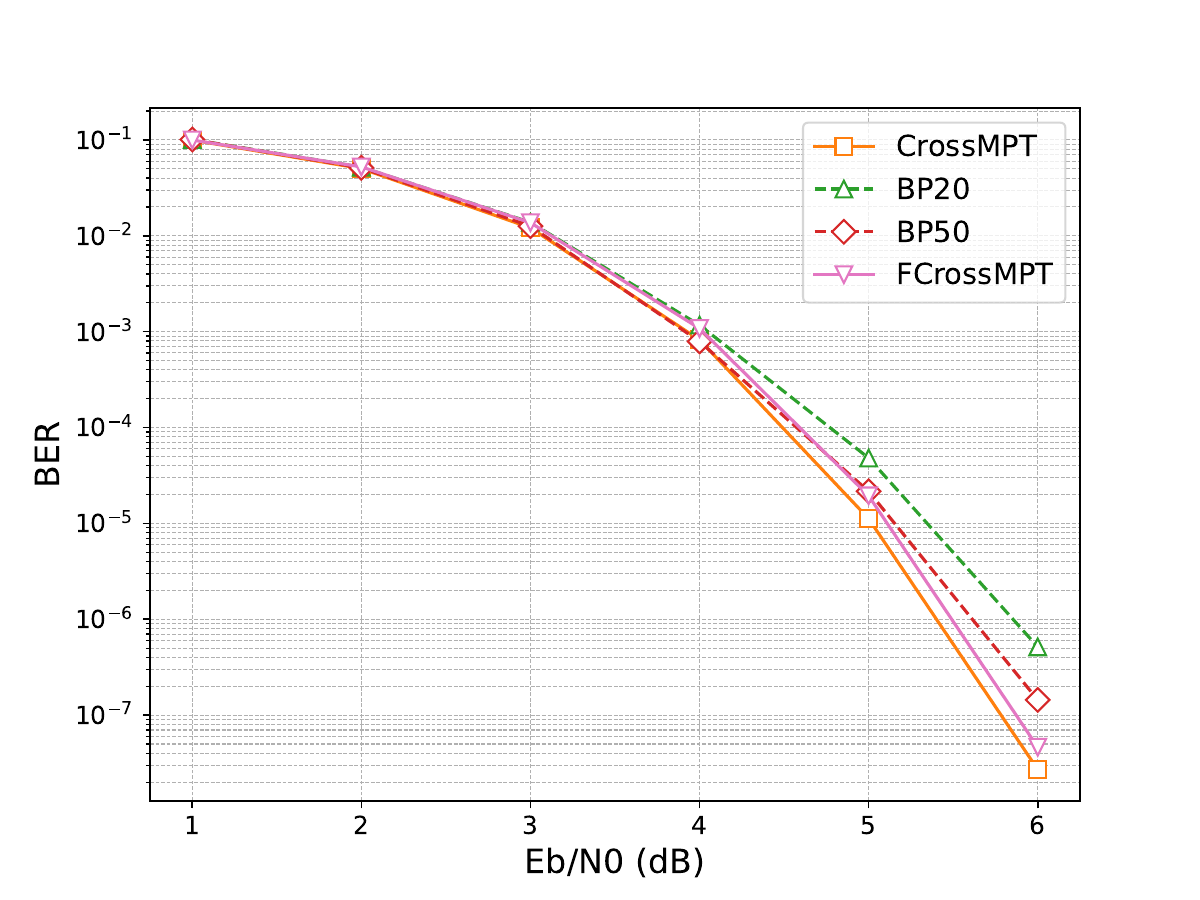}}
\subfloat[\label{fig_121_80_FCross}]{\includegraphics[width=.45\textwidth]{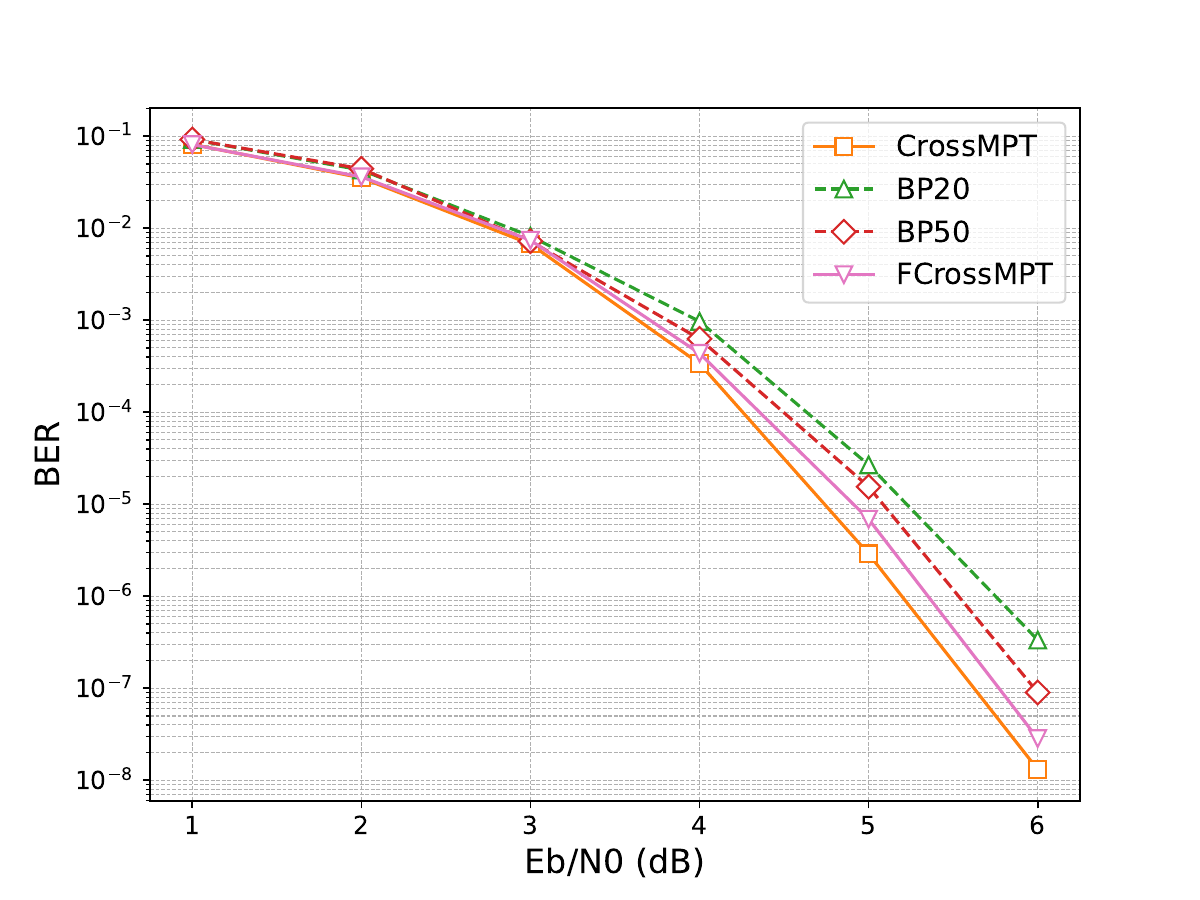}}
\caption{BER performances of CrossMPT, BP decoder~(iteration 20, 50), and FCrossMPT for (a) $(121,70)$ LDPC code and (b) $(121,80)$ LDPC code.
\label{fig_FCross_LDPC}}
\end{center}
\end{figure*}

\begin{figure*}[!t]
\begin{center}
\subfloat[\label{fig_63_45_F}]{\includegraphics[width=.45\textwidth]{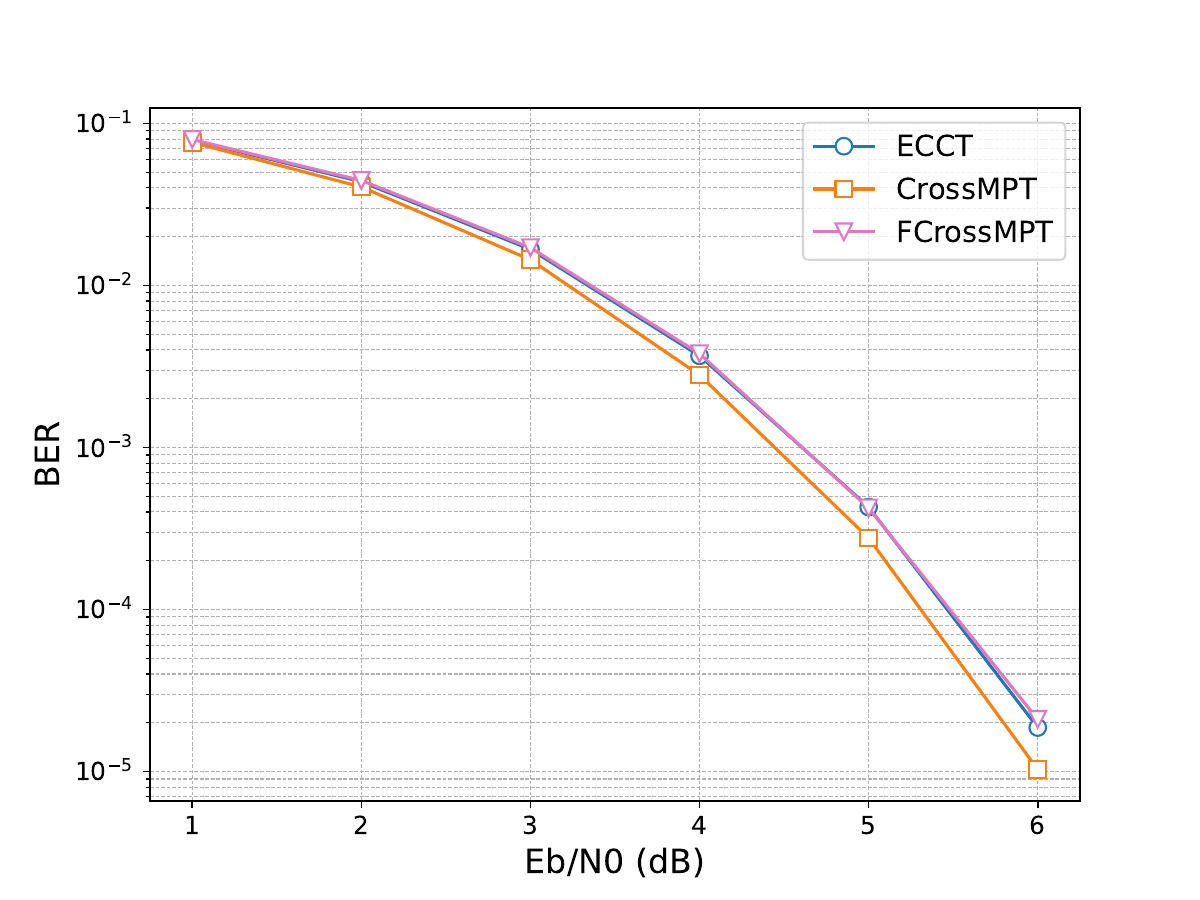}}
\subfloat[\label{fig_49_24_F}]{\includegraphics[width=.45\textwidth]{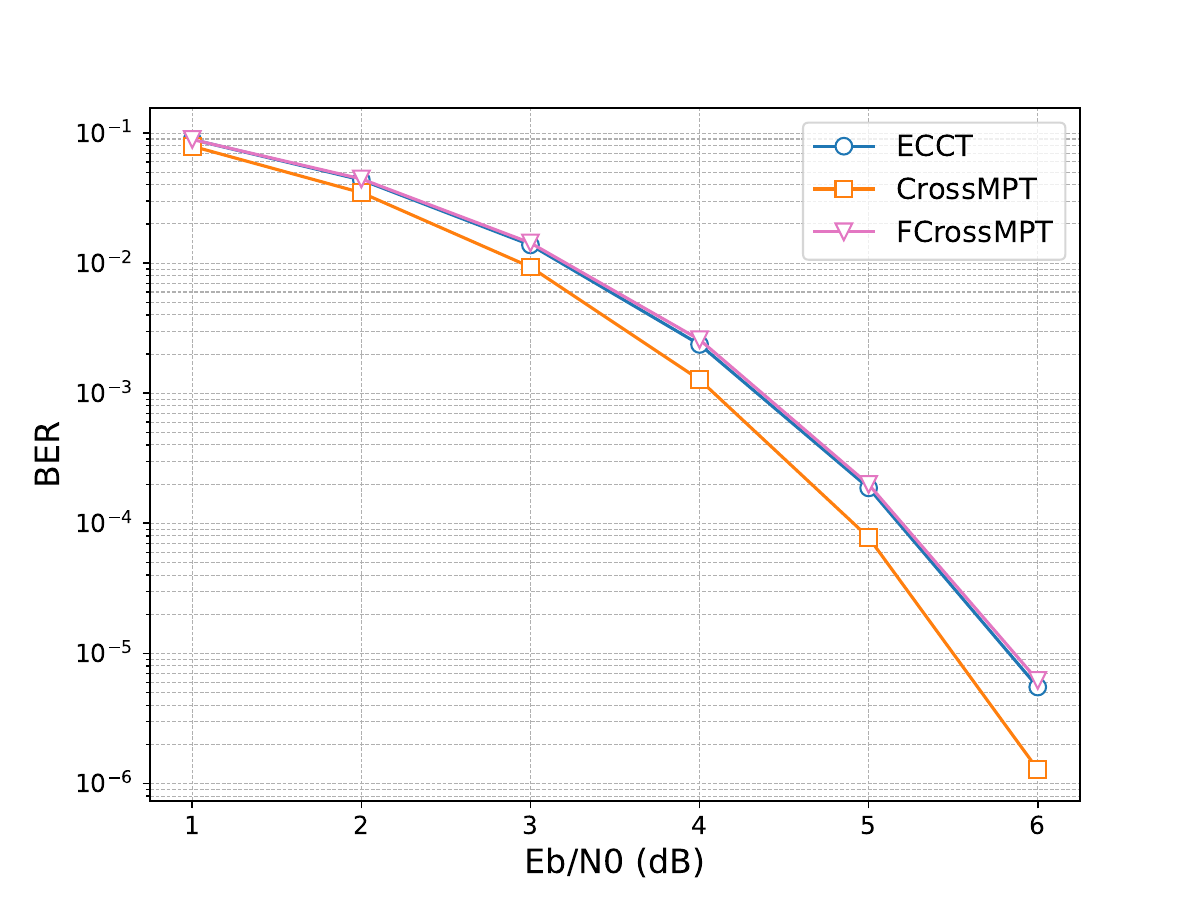}}
\caption{BER performances of ECCT, CrossMPT, and FCrossMPT for (a) $(63,45)$ BCH code and (b) $(49,24)$ LDPC code.
\label{fig_FCrossMPT}}
\end{center}
\end{figure*}

\subsection{Decoding Performance of Code-Agnostic Decoders}

To demonstrate that the proposed FCrossMPT can serve as a unified and code-agnostic decoder suitable for 6G networks, we design two experimental scenarios.
First, FCrossMPT is trained on codes with the same length, but different code rates.
This setup evaluates the model’s ability to generalize across codes with varying code rates within the same code class and length without altering or retraining the model weights.
Specifically, we train the model on LDPC codes of length $n = 121$ but with different code rates: $(121,60)$, $(121,70)$, and $(121,80)$ LDPC codes.

The training setting is identical to the setup in Section~\ref{sec_training}, but the epoch is extended to 3000 due to the diversity of the training data set.
While increasing the number of epochs may further enhance performance, the current epoch value is sufficient to demonstrate the effectiveness of the proposed foundation decoder.
Figs.~\ref{fig_FCross_LDPC}(a) and \ref{fig_FCross_LDPC}(b) compares the decoding performance of BP decoders, CrossMPT, and FCrossMPT for $(121,70)$ and $(121,80)$ LDPC codes.
These codes are included in the training data set, and decoding is performed using a single trained model without any modification to the model weights or structure.
The results confirm that the FCrossMPT achieves a strong decoding performance across all code rates, comparable to CrossMPT, which is trained separately for each codes.
Additionally, the performance of FCrossMPT outperforms the conventional BP decoder, highlighting the strong potential of transformer-based decoders as foundation models.

As the second scenario, we further evaluate the generalization capability of the proposed FCrossMPT by training a single model on a heterogeneous mixture of codes that vary in class, length, and rate.
The training set includes four different codes: $(63,30)$ BCH code, $(63,45)$ BCH code, $(49,24)$ LDPC code, and $(121,60)$ LDPC code using a batch size of 256 over 4000 epochs.
Despite the structural heterogeneity among these codes, the unified model consistently achieves strong decoding performance across all cases as shown in Fig.~\ref{fig_FCrossMPT}.
Although its performance is slightly lower than that of CrossMPT trained on individual codes, it remains comparable to ECCT and demonstrates robust decoding across various code types. These results confirm that FCrossMPT functions effectively as a foundation decoder--even under highly diverse conditions--and highlight its practicality for AI-native, code-agnostic decoding in 6G networks.

\begin{figure*}[!t]
\begin{center}
\subfloat[\label{fig_63_30_BP_DM}]{\includegraphics[width=.45\textwidth]{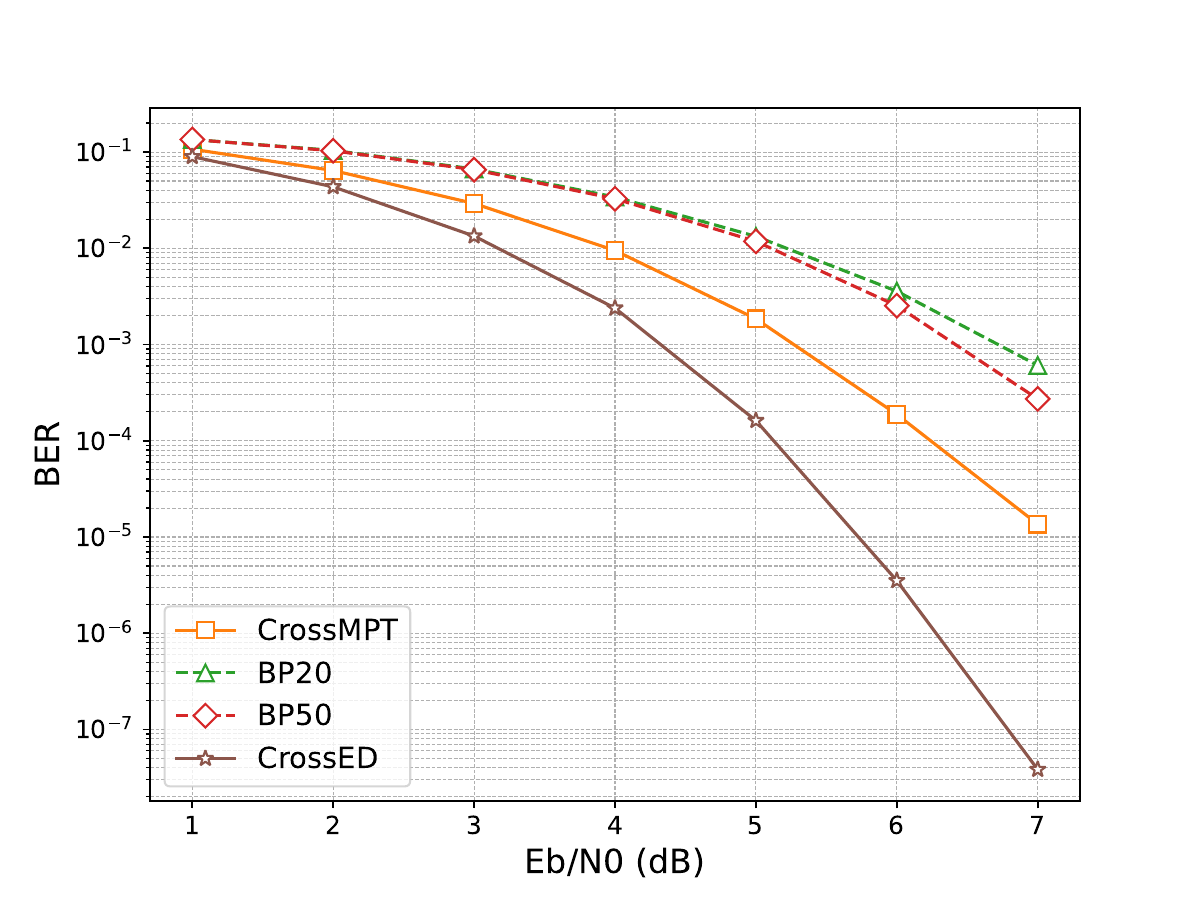}}
\subfloat[\label{fig_127_64_BP_DM}]{\includegraphics[width=.45\textwidth]{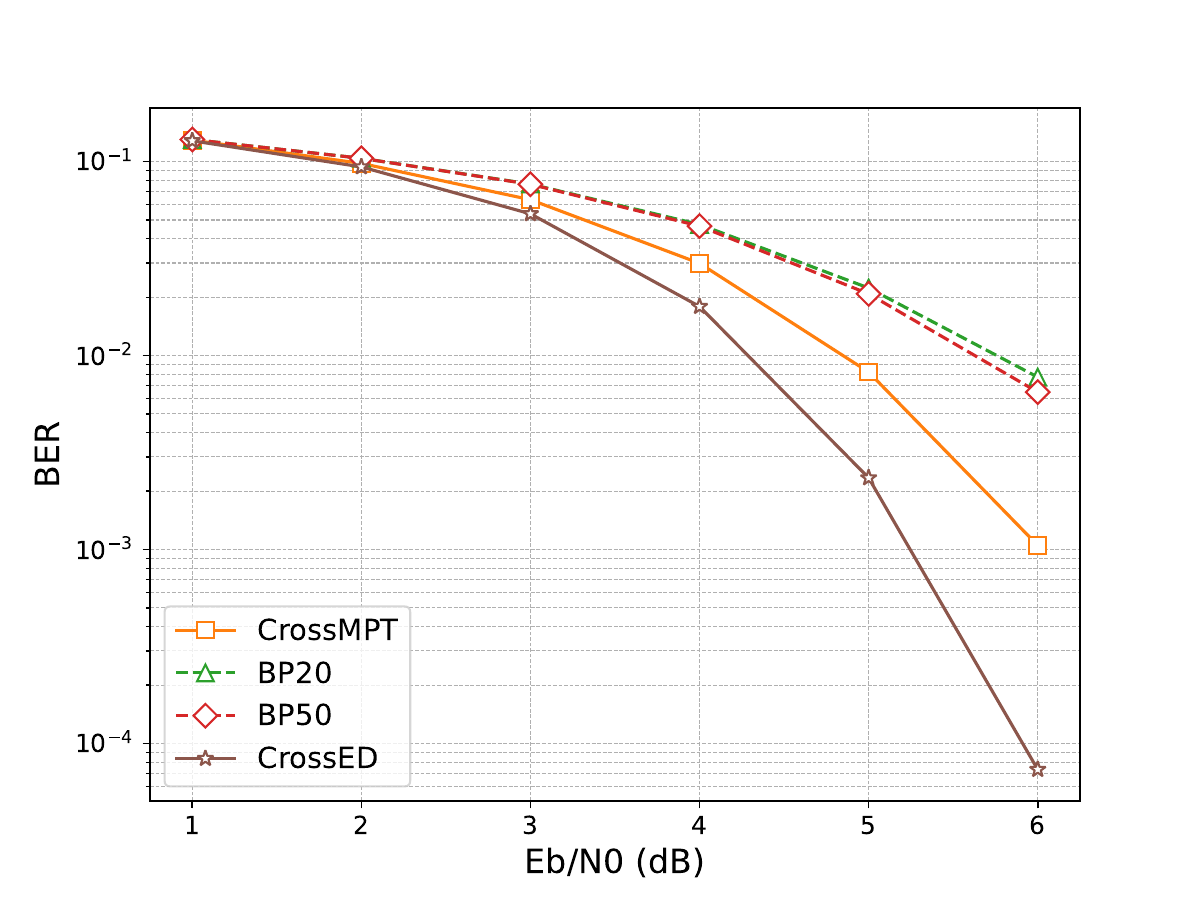}}
\caption{BER performances of CrossMPT, BP decoder~(iteration 20, 50), and CrossED for (a) $(63,30)$ BCH code and (b) $(127,64)$ BCH code.
\label{fig_CrossED_BP_DM}}
\end{center}
\end{figure*}

\begin{figure*}[!t]
\begin{center}
\subfloat[\label{fig_31_16_FCrossED}]{\includegraphics[width=.45\textwidth]{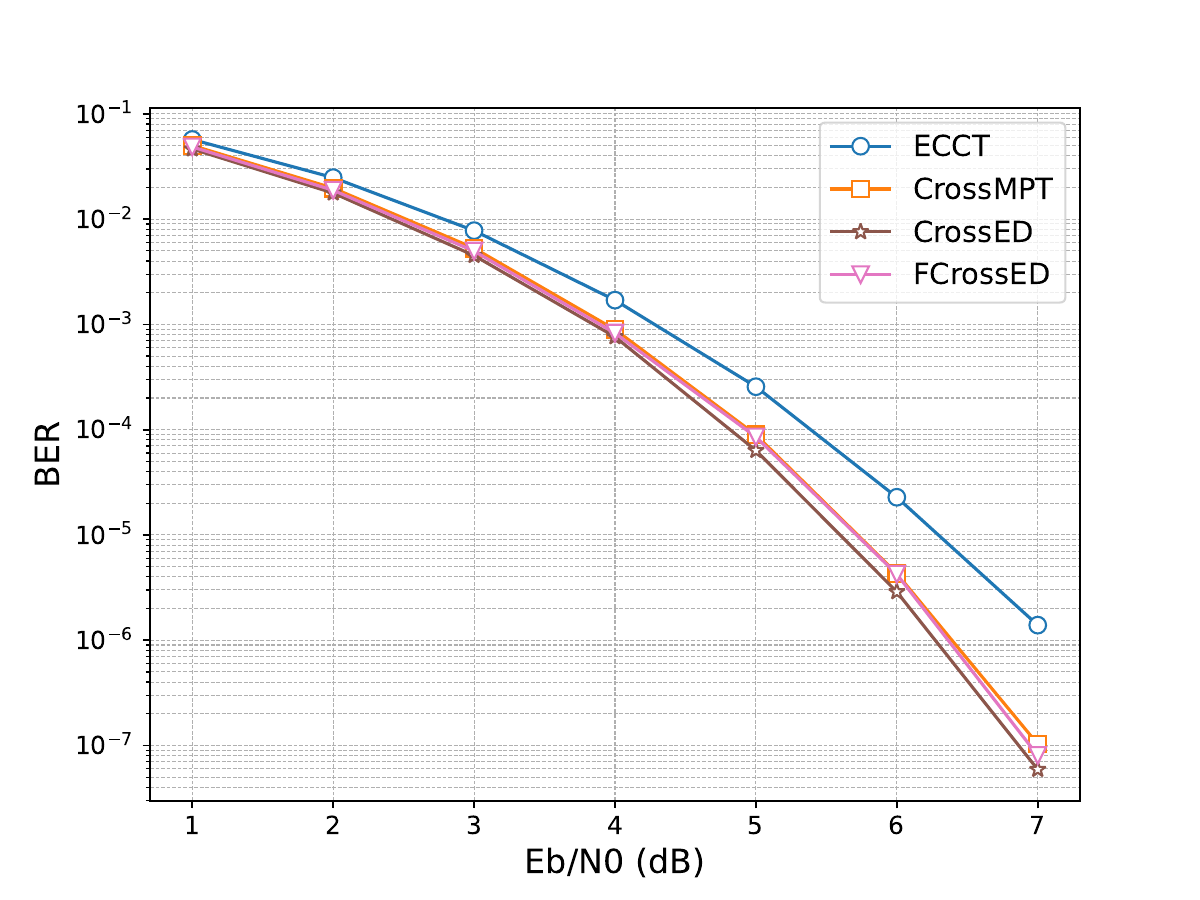}}
\subfloat[\label{fig_63_30_FCrossED}]{\includegraphics[width=.45\textwidth]{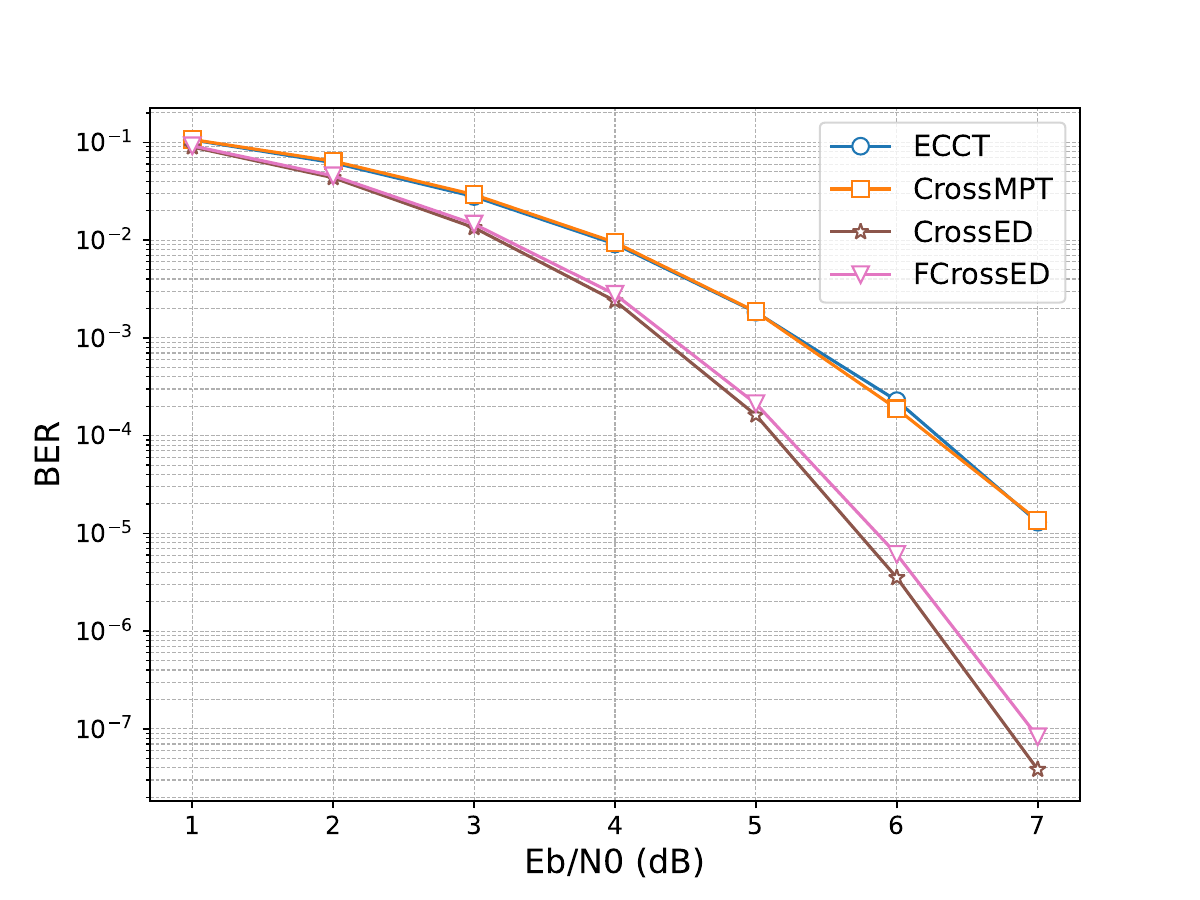}}
\caption{BER performances of ECCT, CrossMPT, CrossED, and FCrossED for (a) $(31,16)$ BCH code and (b) $(63,30)$ BCH code.
\label{fig_FCrossED}}
\end{center}
\end{figure*}

\subsection{Decoding Performance of Ensemble Decoders}

To evaluate the effectiveness of our proposed ensemble decoder, CrossED, we conduct experiments using $p=2$ parallel CrossMPT blocks for the $(63,30)$ BCH and $(127,64)$ BCH codes.
To ensure that most codeword bit positions are covered by the identity matrices in the systematic and complementary PCMs, we focus on codes with a rate close to $1/p$.
The design of CrossED builds upon principles first introduced in MM ECCT~\cite{Park2024multiple}, which demonstrated that using an ensemble of different PCMs can significantly boost the performance of transformer-based decoders.
By leveraging the structural properties of cyclic codes, we construct complementary PCMs by cyclically shifting the columns of the systematic PCM.
This approach enables efficient generation of diverse yet valid PCMs that complement each other in covering different positions.

Fig.~\ref{fig_CrossED_BP_DM} shows the BER performance for CrossED against several baselines.
In Figs.~\ref{fig_CrossED_BP_DM}(a) and \ref{fig_CrossED_BP_DM}(b), CrossED substantial gain over the baseline CrossMPT by more than 2 order in terms of BER for $(63,30)$ BCH code and more than 1 order for $(127,64)$ BCH code.

Notably, CrossED achieves this performance gain without increasing the number of trainable parameters, as all $
p$ CrossMPT blocks share the same weights, and without introducing additional latency, as its ensemble blocks operate in parallel.
Furthermore, the substantial performance improvement in short block length regimes aligns with the goals of next-generation communication systems such as 6G networks, which demand high reliability and low latency communication in short-packet scenarios.
Our findings demonstrate that CrossED offers an effective and efficient solution for decoding short-length codes in future wireless systems.

\subsection{Decoding Performance of Foundation CrossED}

To evaluate CrossED's capability as a foundation decoder, we train CrossED on a dataset comprising three distinct short-length cyclic codes with varying code lengths and code rates: the $(15,7)$ Hamming code, the $(31,16)$ BCH code, and the $(63,30)$ BCH code.
All codes are jointly trained using a batch size of 256 over 5000 epochs.
Figs.~\ref{fig_31_16_FCrossED} and \ref{fig_63_30_FCrossED} compare the decoding performance between ECCT, CrossMPT, CrossED, and FCrossED.
For both the $(31,16)$ and $(63,30)$ BCH codes, FCrossED achieves decoding performance comparable to that of CrossED trained individually on each code.
In other words, CrossED exhibits superior decoding performance while maintaining code-agnostic properties.

\section{Ablation Studies and Analysis}

\begin{figure*}[!t]
\begin{center}
\subfloat[Attention scores with a single bit error \label{fig_attn_score_a}]{\includegraphics[width=\columnwidth]{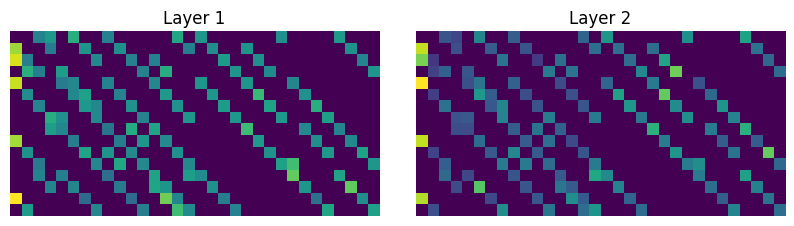}}
\hspace{1cm}
\subfloat[Summation of attention scores with a single bit error \label{fig_attn_score_b}]{\includegraphics[width=.5\columnwidth]{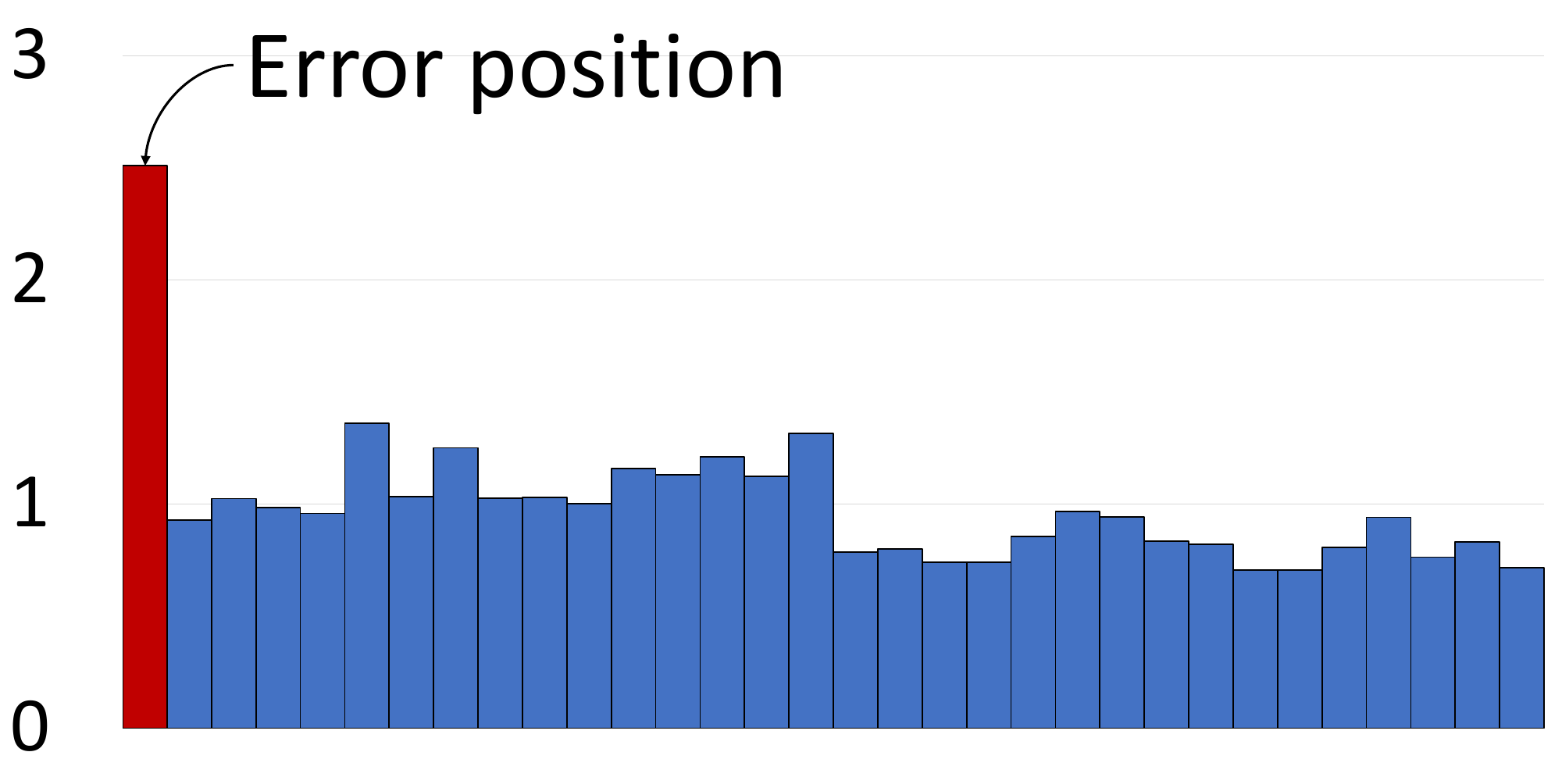}}
\vfill
\subfloat[Attention scores without an error \label{fig_attn_score_c}]{\includegraphics[width=\columnwidth]{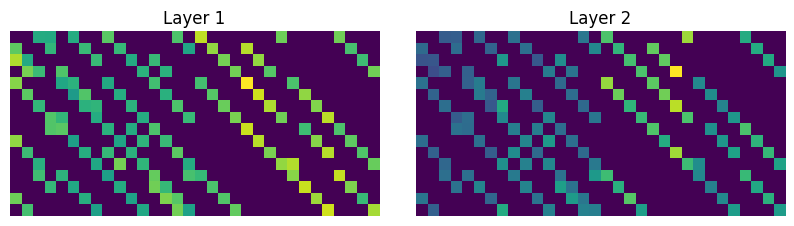}}
\hspace{1cm}
\subfloat[Summation of attention scores without an error \label{fig_attn_score_d}]{\includegraphics[width=.5\columnwidth]{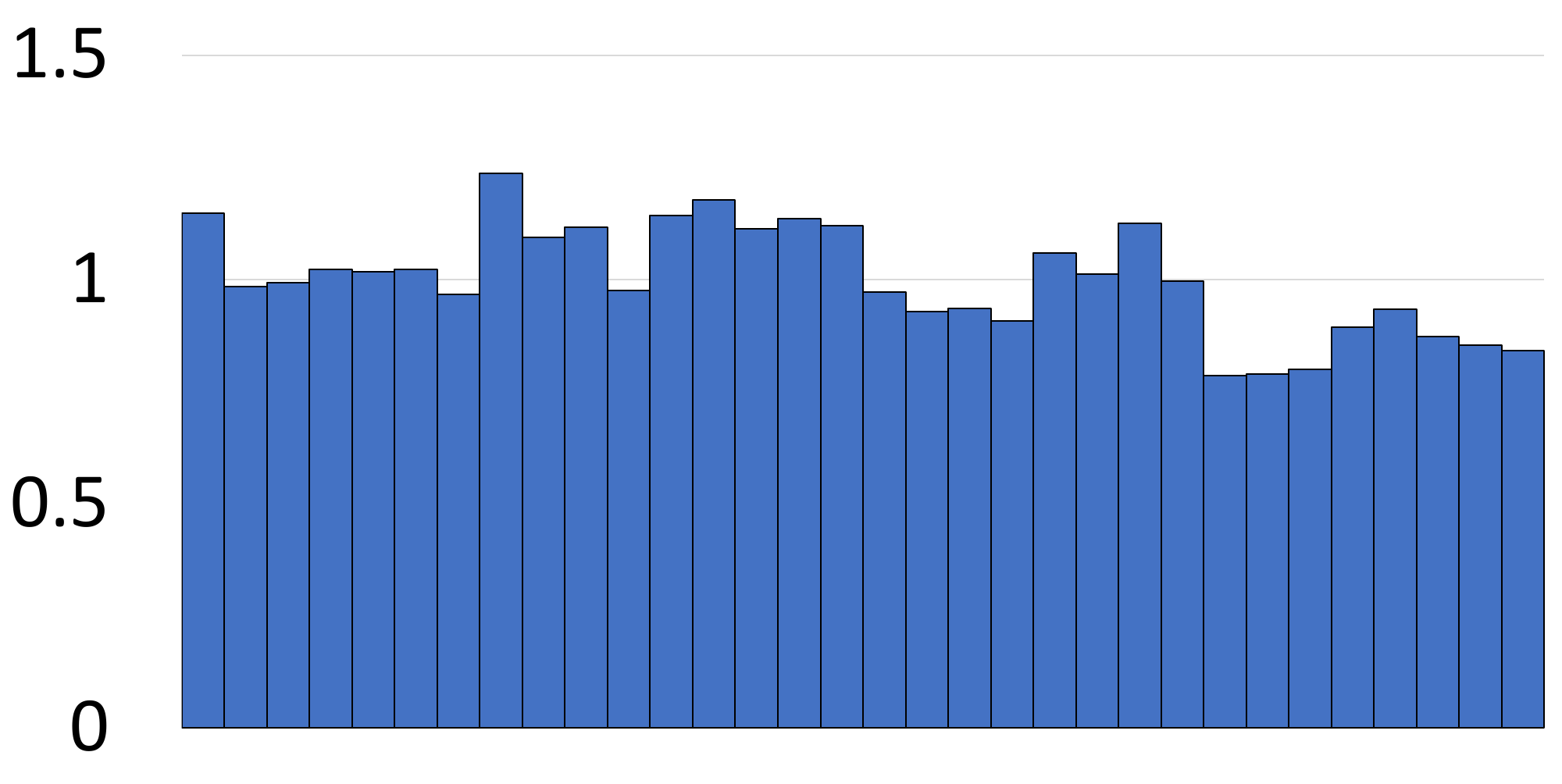}}
\caption{(a) Attention scores when a single-bit error occurs at the first bit position; (b) vertical-wise sum of attention scores in (a); (c) attention scores in the absence of errors; (d) vertical-wise sum of attention scores in (c).
\label{fig_attn_score}}
\end{center}
\vspace{-4mm}
\end{figure*}

\subsection{Visualization of Cross-Attention Map}

To understand the internal operation of CrossMPT, we visualize the attention maps in Fig.~\ref{fig_attn_score}(a) by intentionally corrupting a specific bit of the $(32,16)$ LDPC code and analyzing the resulting attention distributions.
Fig.~\ref{fig_attn_score}(a) shows the attention scores for the first two decoding layers, along with the sum of these scores in Fig.~\ref{fig_attn_score}(b).
The summation is carried out vertically to highlight the overall attention score for each bit.
As shown in Fig.~\ref{fig_attn_score}(b), the attention score of the first bit~(or first column) exhibits a noticeably higher attention score compared to the others, indicating that CrossMPT correctly identifies and emphasizes the erroneous bit.
However, once the error is corrected, this emphasis disappears, and CrossMPT no longer assigns disproportionately high attention to that bit position.
Figs.~\ref{fig_attn_score}(c) and \ref{fig_attn_score}(d) show the attention maps and their summations when no errors are present.
In this case, the attention scores are more evenly distributed across all bit positions, confirming that CrossMPT selectively allocates attention only when needed.

\begin{table}[!t]
\caption{Comparison of decoding performance for ECCT with additional masking, standard ECCT, and CrossMPT. The results are measured by the negative natural logarithm of BER.
The best results are highlighted in \textbf{bold}.
Higher is better.\label{tab_appendix_ablation_mask}}
\begin{center}
\resizebox{\columnwidth}{!}{\begin{tabular}{cccccccccc}
\toprule
\multicolumn{1}{c}{Method} & \multicolumn{3}{c}{ECCT + Masking} & \multicolumn{3}{c}{ECCT} & \multicolumn{3}{c}{CrossMPT}\\
\cmidrule(r){1-1}\cmidrule(r){2-4}\cmidrule(r){5-7}\cmidrule(r){8-10}
\multicolumn{1}{c}{Parameter}    & 4   & 5    & 6    & 4   & 5    & 6    & 4   & 5    & 6\\
\midrule
$(31,16)$ BCH   & 6.52 & 8.55 & 11.42 & 6.39 & 8.29 & 10.66 & \textbf{6.98} & \textbf{9.25} & \textbf{12.48} \\
$(63,45)$ BCH   & 5.53 & 7.74 & 10.88 & 5.60 & 7.79 & 10.93 & \textbf{5.90} & \textbf{8.20} & \textbf{11.62} \\
$(64,48)$ Polar   & 6.25 & 8.26 & 10.93 & 6.36 & 8.46 & 11.09 & \textbf{6.51} & \textbf{8.70} & \textbf{11.31} \\
$(121,60)$ LDPC  & 4.98 & 7.91 & 12.61 & 5.17 & 8.31 & 13.30 & \textbf{5.74} & \textbf{9.26} & \textbf{14.78} \\
\bottomrule
\end{tabular}}
\end{center}
\end{table}

\subsection{Ablation Studies with Additional Masking in ECCT}

\begin{figure}[!t]
\begin{center}
\centerline{\includegraphics[width=.9\columnwidth]{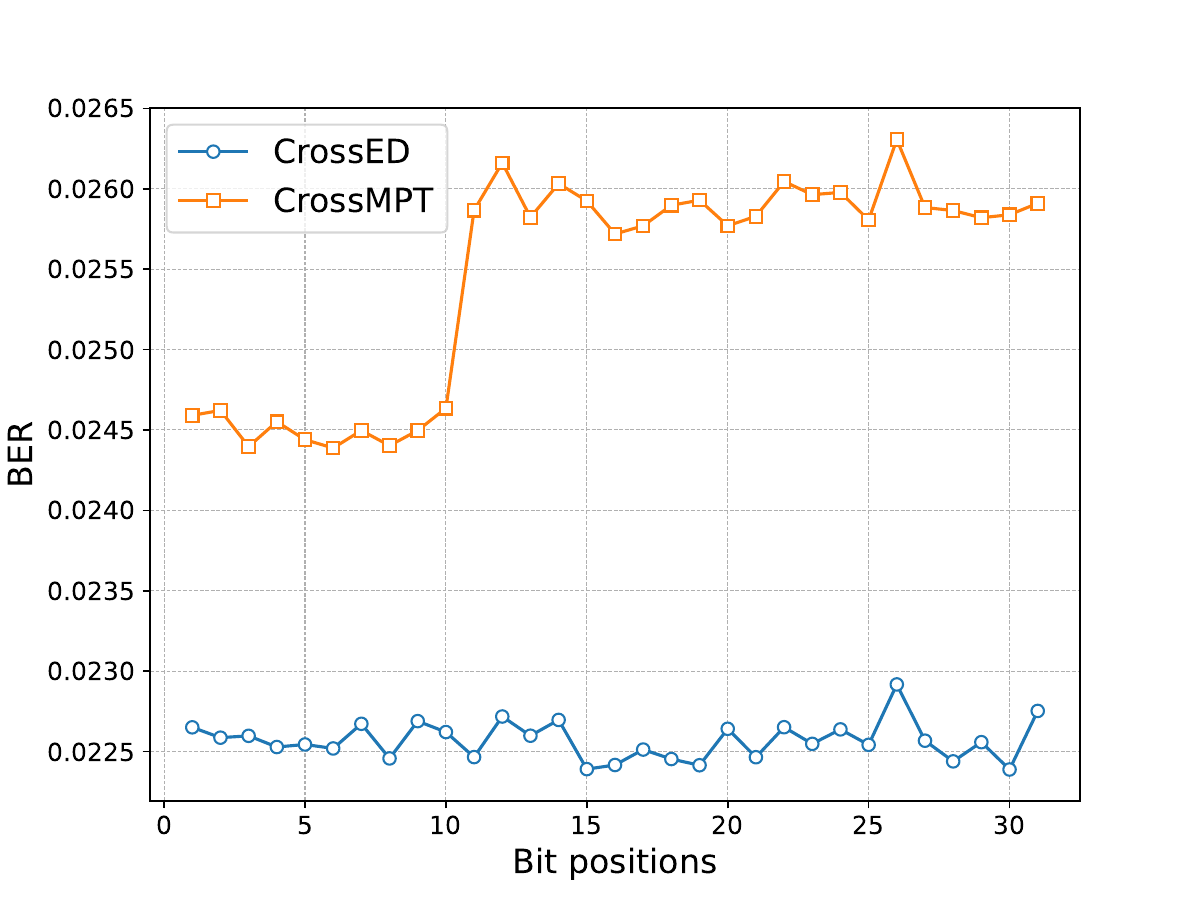}}
\caption{Bitwise BER performance of CrossMPT and CrossED for $(31,21)$ BCH code.\label{fig_bitwise}}
\end{center}
\vspace{-6mm}
\end{figure}

CrossMPT demonstrates that focusing on magnitude–syndrome relationships alone is sufficient to achieve high decoding performance.
To assess the contribution of magnitude–magnitude and syndrome–syndrome relationships in the conventional ECCT architecture, we evaluate the decoding performance of ECCT when additional masking is applied to all positions corresponding to these two types of relationships.
Table~\ref{tab_appendix_ablation_mask} compares the decoding performance of the additionally masked ECCT, standard ECCT, and CrossMPT.
The results show no significant performance degradation in the additionally masked ECCT, indicating that the magnitude-magnitude and syndrome-syndrome relationships are not critical for decoding.
This implies that the conventional ECCT could be enhanced by focusing on the more critical relationships, as CrossMPT achieves this by explicitly eliminating low-attention interactions and concentrating solely on the magnitude–syndrome relationship.


\subsection{Effects of Ensemble Decoder}

Fig.~\ref{fig_bitwise} compares the bitwise BER performances of CrossMPT and CrossED~($p=3$) for the $(31,21)$ BCH code.
CrossMPT employs the systematic PCM as a mask matrix, where the left side of the PCM contains an identity matrix.
As a result, positions corresponding to this identity matrix exhibit notably lower BER, benefiting from stronger error-correcting capabilities.
In contrast, CrossED utilizes multiple complementary PCMs, including the systematic PCM and two additional complementary PCMs derived by cyclically shifting the identity matrix.
This approach provides a more strong error correcting capability across all bit positions, effectively improve the decoding performance compared to CrossMPT.
This result demonstrates that even in transformer-based decoders, employing ensemble structure with multiple PCMs significantly enhances decoding performance without increasing latency.
This approach aligns well with 6G network requirements.


\begin{figure*}[!t]
\begin{center}
\centerline{\includegraphics[width=.8\textwidth]{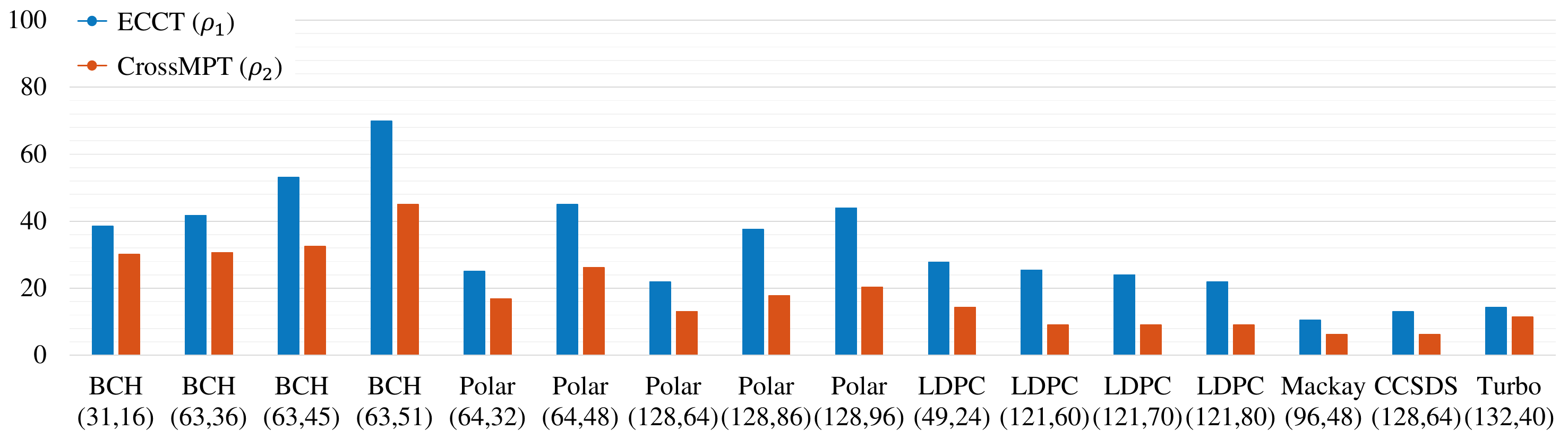}}
\caption{Comparison of the mask matrix density between ECCT and CrossMPT.\label{fig_sparsity}}
\end{center}
\vspace{-6mm}
\end{figure*}

\begin{table*}[!t]
\caption{Comparison of FLOPs, inference time, and training time between ECCT and CrossMPT for various codes.
Inference time is measured for decoding a single codeword and training time is measured for a single epoch.\label{tab_time}}
\begin{center}
\resizebox{.95\textwidth}{!}{\begin{tabular}{cccccccccccc}
\toprule
\multirow{2}{*}{Codes} & \multirow{2}{*}{Parameter} & \multicolumn{2}{c}{FLOPs} & \multicolumn{2}{c}{Inference (codeword)} & \multicolumn{2}{c}{Training (epoch)} & \multicolumn{2}{c}{Mask density} & \multicolumn{2}{c}{Memory usage}\\
\cmidrule(r){3-4}\cmidrule(r){5-6}\cmidrule(r){7-8}\cmidrule(r){9-10}\cmidrule(r){11-12}
                                &                            & CrossMPT  & ECCT     & CrossMPT  & ECCT     & CrossMPT  & ECCT  & CrossMPT  & ECCT & CrossMPT  & ECCT  \\ 
\midrule
\multirow{1}{*}{BCH}   & (63,45)                             & 99.8~M               & 106.4~M              & 326~$\mu$s                & 328~$\mu$s              & 29~s                 & 29~s              & 32.45\%               & 53.09\%  & 962 MiB               & 1828 MiB\\\midrule 
\multirow{2}{*}{LDPC}  & (121,70)                            & 229.7~M               & 256.8~M              & 400~$\mu$s                & 450~$\mu$s              & 58~s                 & 80~s              & 9.09\%                & 24.01\%  & 1980 MiB               & 3926 MiB\\ 
                       & (121,80)                            & 212.5~M               & 238.0~M              & 391~$\mu$s                & 436~$\mu$s              & 53~s                 & 76~s              & 9.09\%                & 21.94\%  & 1936 MiB               & 3602 MiB\\\midrule 
\multirow{1}{*}{Turbo} & (132,40)                            & 303.6~M               & 343.4~M              & 459~$\mu$s                & 511~$\mu$s              & 83~s                 & 110~s             & 11.43\%               & 14.25\% & 2362 MiB               & 5580 MiB\\\midrule 
\multirow{1}{*}{BCH}  & (255,223)                           & 28.2~M               & 53.5~M              & 747~$\mu$s               & 859~$\mu$s             & 56~s                & 145~s             & 48.63\%                & 78.21\% & 1036 MiB               & 7318 MiB\\\midrule 
\multirow{1}{*}{WRAN}  & (384,320)                           & 53.1~M               & 111.3~M              & 1295~$\mu$s               & 1638~$\mu$s             & 104~s                & 305~s             & 5.21\%                & 13.25\% & 3270 MiB               & 18192 MiB\\ 
\bottomrule
\end{tabular}}
\end{center}
\vspace{-4mm}
\end{table*}

\subsection{Complexity Analysis}

To maintain parameter efficiency, the two cross-attention blocks within each CrossMPT decoder layer utilize shared parameters. Specifically, the weight matrices ($\mathbf{W}_{\rm Q}, \mathbf{W}_{\rm K}, \mathbf{W}_{\rm V}$), normalization layers, and FFNN layers all share parameters because training them separately yields negligible performance differences.
As a result, CrossMPT has the same number of trainable parameters as the original ECCT.

A key architectural advantage of CrossMPT is its sparser attention mechanism compared to ECCT, as illustrated by their respective mask matrices in Fig.~\ref{fig_mask}.
The mask matrix for the original ECCT, shown in Fig.~\ref{fig_mask}, includes a dense $n \times n$ submatrix (depicted in white) representing depth-2 Tanner graph connections~\cite{b_ECCT}, which increases the number of unmasked positions and the computational complexity.
In contrast, the mask matrices for CrossMPT are predominantly shown in blue, indicating that their attention matrices are sparser.
This difference in density is quantified in Fig.~\ref{fig_sparsity}.
For all codes, the mask matrix of CrossMPT is sparser than ECCT, implying that CrossMPT can achieve lower computational complexity.

For ECCT, the complexity of the self-attention mechanism without masking is $\mathcal{O}(N(d^2(2n-k)+(2n-k)^2d))$.
With masking, this reduces to $\mathcal{O}(N(d^2(2n-k)+hd))$~\cite{b_ECCT}, where $h = \rho_{1}(2n-k)^2$ denotes the fixed number of computations of the self-attention module and $\rho_{1}$ denotes the density of the mask matrix in ECCT.
In comparison, the two cross-attention modules of CrossMPT have a complexity of $\mathcal{O}(N(d^2(2n-k)+2n(n-k)d))$ before masking.
When masking is taken into account, the complexity can be reduced to $\mathcal{O}(N(d^2(2n-k)+(2\tilde{h})d))$, where $\tilde{h} = \rho_{2} n(n-k)$ denotes the number of computations of a single cross-attention module and $\rho_{2}$ denotes the density of the mask matrix in CrossMPT.
As established in Fig.~\ref{fig_sparsity}, $\rho_1 > \rho_2$, which leads to the conclusion that $h > 2\tilde{h}$.
This confirms that CrossMPT achieves a formal reduction in computational complexity compared to the original ECCT.
This reduced complexity during the training and testing are detailed in Table~\ref{tab_time}, which compares the total FLOPs, inference time, training time, and memory usage of ECCT and CrossMPT.
All results are obtained for $N=6$ and $d=128$, except for $(255,223)$~BCH code and $(384,320)$ WRAN LDPC code, which are obtained for $N=6$ and $d=32$.
The inference time refers to the duration required to decode a single codeword and the training time measures the duration to complete one epoch of training.
The results show that CrossMPT consistently outperforms ECCT across these metrics.
The reduction in FLOPs directly translates to shorter inference and training durations. Furthermore, CrossMPT offers a significant advantage in memory usage, particularly for long codes. This memory efficiency stems from the smaller attention map size required by CrossMPT's architecture, which is $2n(n-k)$ compared to $(2n-k)^2$ for ECCT.

While FCrossMPT is not explicitly included in Table~\ref{tab_time}, its computational complexity and memory usage are expected to be nearly identical to those of CrossMPT.
This is because the core architecture, particularly the $N$ CrossMPT blocks remain unchanged.
The primary modifications in FCrossMPT are in the initial embedding and final output layers, where code-dependent parameters are replaced with code-invariant ones.
These changes have a negligible impact on the overall FLOPs and inference time compared to the main decoding loop, while significantly enhancing the model's generalization capabilities.

As shown in Fig.~\ref{fig_CrossED}, the decoder layer in CrossED requires $p$ times more FLOPs than that of CrossMPT.
However, since all CrossMPT blocks share the same parameters, the number of parameters remains the same as in CrossMPT.
However, in terms of decoding latency, CrossED remains comparable to the case of $p=1$ due to its parallel processing of the $p$ self-attention modules.

\section{Conclusion}

In this paper, we proposed an AI-native code-agnostic decoder based on transformer architectures to support the flexibility, scalability, and generalization requirements of 6G networks.
We first introduced CrossMPT, a message-passing-like decoder that employs masked cross-attention modules to iteratively update the input magnitude and syndrome representations. CrossMPT outperforms conventional transformer-based decoders while reducing decoding complexity.
To move toward a code-agnostic foundation model, we developed FCrossMPT by replacing code-dependent parameters with code-invariant embeddings.
To move toward a code-agnostic foundation model, we developed FCrossMPT by replacing code-dependent parameters with code-invariant embeddings.
Moreover, we proposed CrossED, an ensemble decoder consisting of $p$ parallel CrossMPT blocks, each using a different PCM.
CrossED further improves the decoding performance for short blocklength codes and also serves as a foundation model, referred to as FCrossED.
Our extensive experiments demonstrate that CrossMPT and CrossED achieve state-of-the-art decoding performance, while FCrossMPT and FCrossED exhibit excellent performance as foundation models, making them strong candidates for unified, code-agnostic decoders in 6G networks.

    \appendices
	
    \bibliographystyle{IEEEtran}
    \bibliography{abrv,mybib}

\end{document}